\documentclass[aps,twocolumn]{revtex4-1}
\usepackage{amsmath}
\usepackage{color}
\usepackage{xcolor}
\usepackage{subfigure}
\usepackage[outercaption,wide]{sidecap}
\usepackage{graphicx}
\usepackage{tikz, tikzscale}

\usetikzlibrary{calc,patterns,angles,quotes, arrows.meta, decorations.pathmorphing, math, decorations.pathreplacing}

\newcommand \rmu{{{\rm u}}}

\newcommand \tL{{{\tilde{L}}}}

\newcommand \tT{{{\tilde{T}}}}
\newcommand \tDelta{{{\tilde{\Delta}}}}

\newcommand \bfx{{{\bf{x}}}}
\newcommand \syy{{{\sigma_{yy}}}}

\newcommand \sxx{{{\sigma_{xx}}}}

\newcommand \Keff{K_{\rm eff}}
\newcommand \Ksub{K_{\rm sub}}
\newcommand \Kcurv{K_{\rm curv}}
\newcommand \Ktens{K_{\rm tens}}
\newcommand \bKtens{\bar{K}_{\rm tens}}

\newcommand \beq{\begin{equation}}
\newcommand \eeq{\end{equation}}

\newcommand \UTFT{U_{\rm TFT}}

\newcommand \Usub{U_{\rm sub}}
\newcommand \tDeltaTFT{{{\tilde{\Delta}}^{TFT}}}
\newcommand \PhiTFT{{{\Phi}^{TFT}}}

\DeclareFontFamily{OMS}{oasy}{\skewchar\font48 }
\DeclareFontShape{OMS}{oasy}{m}{n}{%
         <-5.5> oasy5     <5.5-6.5> oasy6
      <6.5-7.5> oasy7     <7.5-8.5> oasy8
      <8.5-9.5> oasy9     <9.5->  oasy10
      }{}
\DeclareFontShape{OMS}{oasy}{b}{n}{%
       <-6> oabsy5
      <6-8> oabsy7
      <8->  oabsy10
      }{}
\DeclareSymbolFont{oasy}{OMS}{oasy}{m}{n}
\SetSymbolFont{oasy}{bold}{OMS}{oasy}{b}{n}

\DeclareMathSymbol{\smallleftarrow}     {\mathrel}{oasy}{"20}
\DeclareMathSymbol{\smallrightarrow}    {\mathrel}{oasy}{"21}
\DeclareMathSymbol{\smallleftrightarrow}{\mathrel}{oasy}{"24}

\begin{document}

\title{Stretching Hookean ribbons: from buckling instability to tensional wrinkling}  
\title{Stretching Hookean ribbons \\
Part II: from buckling instability to far-from-threshold wrinkle pattern} 
%
\author{Meng Xin and Benny Davidovitch}
\affiliation{Physics Department, University of Massachusetts, Amherst MA 01003}
\begin{abstract}
We address the fully-developed wrinkle pattern formed upon stretching a Hookean, rectangular-shaped sheet, 
when the longitudinal tensile load induces transverse compression that 
far exceeds the stability threshold of a purely planar deformation. 
At this ``far from threshold'' parameter regime, which has been the subject of the celebrated Cerda-Mahadevan (CM) model 
\cite{Cerda03}, the wrinkle pattern expands throughout the length of the sheet and the characteristic wavelength of undulations 
is much smaller than its width. Employing Surface Evolver simulations over a range of sheet thicknesses and tensile loads 
we elucidate the theoretical underpinnings of 
the far-from-threshold framework in this set-up. We show that 
the evolution of wrinkles comes in tandem with collapse of transverse compressive stress, rather than vanishing transverse strain (which was hypothesized in \cite{Cerda03}), such that the stress field approaches asymptotically a compression-free limit, describable by tension field theory. 
We compute the compression-free stress field by simulating a Hookean sheet that has finite stretching modulus but no bending rigidity, 
and show that this singular limit encapsulates the geometrical nonlinearity underlying the amplitude-wavelength ratio of wrinkle patterns in physical, highly bendable sheets, 
even though the actual strains may be so small that the local mechanics is perfectly Hookean. 
Finally, we revisit the balance of bending and stretching energies that gives rise to a favorable wrinkle wavelength, 
and study the consequent dependence of the wavelength on the tensile load as well as the thickness and length of the sheet.     
\end{abstract}
\maketitle

\section{Introduction}

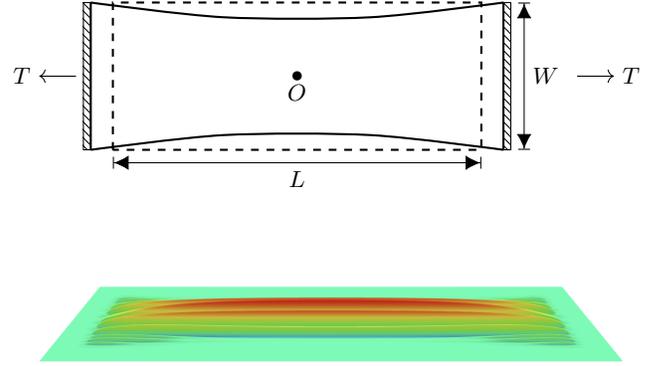
\begin{figure}
\begin{tikzpicture}
    \node at (0,0) 
    {\resizebox{\linewidth}{!}{\begin{tikzpicture}[scale=1]
    \tikzmath{\xshift=0.3; \yshift=0.2; \halfw=1; \halfl=2.5; };
\coordinate (LU) at (-\halfl, \halfw); 
\coordinate (LL) at (-\halfl, -\halfw);
\coordinate (RU) at (\halfl,\halfw);
\coordinate (RL) at (\halfl,-\halfw);
\coordinate (rLU) at ($(LU) + (-\xshift, 0) $); 
\coordinate (rLL) at ($(LL) + (-\xshift, 0) $);
\coordinate (rRU) at ($(RU) + (\xshift, 0) $);
\coordinate (rRL) at ($(RL) + (\xshift, 0) $);
\coordinate (sLU) at ($(-1, \halfw) + (0, -\yshift)$);
\coordinate (sLL) at ($(-1, -\halfw) + (0, \yshift)$);
\coordinate (sRU) at ($(1, \halfw) + (0, -\yshift)$);
\coordinate (sRL) at ($(1, -\halfw) + (0, \yshift)$);
\draw[dashed, thick] (LL) rectangle (RU);
\draw[thick] (rLU) -- (rLL);
\draw[thick] (rRU) -- (rRL);
\draw[thick] plot [smooth, tension=0.5] coordinates {(rLU) (sLU) (sRU)  (rRU)};
\draw[thick] plot [smooth, tension=0.5] coordinates {(rLL) (sLL) (sRL)  (rRL)};
\fill[pattern=north west lines] (rLL) rectangle ($(rLU) + (-0.1, 0)$);
\fill[pattern=north west lines] (rRL) rectangle ($(rRU) + (0.1, 0)$);
\draw (rLL) rectangle ($(rLU) + (-0.1, 0)$);
\draw (rRL) rectangle ($(rRU) + (0.1, 0)$);
\draw[->] ({-\halfl -\xshift -0.2}, 0) -- ++(-0.5,0) node[left] {$T$};
\draw[->] ({\halfl +\xshift +1}, 0) -- ++(0.5,0) node[right] {$T$};
\draw[{|Latex[width=2mm,length=2mm]}-{Latex[width=2mm,length=2mm]|}] ([yshift=-5pt]LL) -- ([yshift=-5pt]RL) node[midway, below] {$L$};
\draw[{|Latex[width=2mm,length=2mm]}-{Latex[width=2mm,length=2mm]|}] ([xshift=8pt]rRU) -- ([xshift=8pt]rRL) node[midway, right] {$W$};
\node[right, below] at (0,0) {$O$};
\node at (0,0) {$\bullet$};
\end{tikzpicture}}};
    \node at (0,-3) {\includegraphics[width=0.9\linewidth]{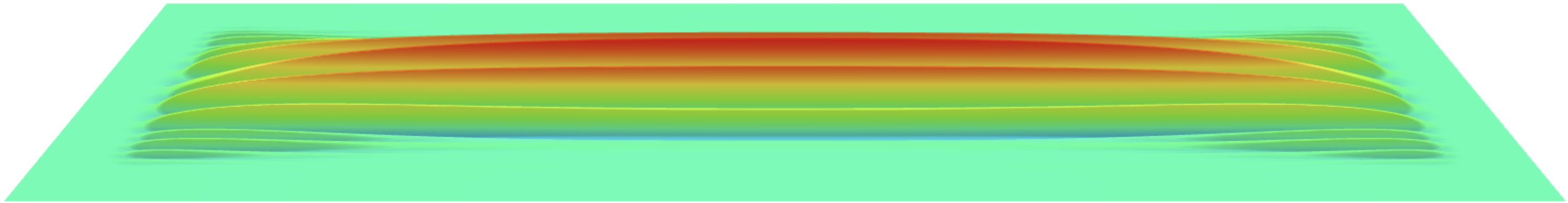}};
\end{tikzpicture}
\caption{
(A) Schematic drawing of a rectangular sheet with width $W$ and length $L=$, subjected to longitudinal tensile loads, $T = \tT \cdot Y $ ($force/length$)  
that pull on the two short edges, $x=\pm L/2$, while the long edges are free. The short edges are clamped, such that both normal (out-of-plane) displacement, $\zeta(x=\pm L/2,y)$, and transverse (in-plane) displacement, $\rmu_y(x=\pm L/2,y)$, 
vanish, 
and their longitudinal displacements are $\rmu_x(x=\pm L/2,y) \approx \pm \tT L/2$. (B) Characteristic wrinkle pattern attained by a highly bendable sheet ($\epsilon \approx  0.0085 \epsilon_c$) in the Hookean FT parameter regime.}
\label{fig:schematic}
\end{figure}

A classic example of 
energy-driven 
pattern formation in materials \cite{Kohn06} 
is the parallel wrinkles 
that emerge in a rectangular solid sheet 
upon clamping its short edges and pulling them apart (Fig.~\ref{fig:schematic}). 
Despite the apparent simplicity of this phenomenon that beautifully demonstrates 
the spontaneous emergence of 
patterns in continuous media emerge through symmetry-breaking instability of a homogeneous, featureless ``base state'',  
the actual mechanism underlying tension-induced wrinkles that permeate most of the stretched sheet is rather subtle and, arguably, counter-intuitive.   
A first difficulty 
pertains to the 
non-intuitive nature of the base state -- a planar deformation of the sheet where 
the exerted  
longitudinal tension $T$ conspires with the clamping of the short edges to induce 
transversely compressed zones, localized  at a short distance from the clamped edges \cite{Friedl2000, Nayya11}.
In a preceding paper \cite{Xin20} we showed that as 
the tensile load exceeds a thickness-dependent threshold $T_c$  
the sheet undergoes an Euler-like instability in the transversely-compressed zones,  
where the ``wavelength'' of the buckled shape is determined solely by the sheet's width $W$, and unaffected  
by its thickness ($t \ll W$) nor by its length ($L \gg W$). 
A second difficulty, which we address in this article, is the transition from the near-threshold pattern of localized buckling 
at $T \gtrsim T_c$ to a 
pattern of fine, elongated wrinkles that pervade 
the stretched sheet, whose characteristic ``wavelength'' $\lambda \ll W$ depends on the sheet's thickness $t$ and the 
tensile load $T$. 

Realizing that the observed wrinkle pattern in this system  
cannot be described  
by a standard ``post-buckling'' approach, 
in which the out-of-plane deflection 
is assumed 
a perturbation of 
the planar state \cite{TimoshenkoBook}, 
numerous researchers employed 
non-Hookean models, attempting to capture the elastic response of the stretched sheet at finite ($O(1)$) strain 
\cite{Nayya11,Healy13,Healy16, Fu19, Sipos16, Nayya14,Wang19, Pontanescu19,Fried12}. However, while certain aspects of this problem do indeed stem from non-Hookean response (most notably, the reentrance of a stable planar state when the exerted tensile strain exceeds a finite value, typically $0.3-0.4$ \cite{Healy13,Healy16,Wang19,Sipos16,Nayya14}, the transition from the near-threshold localized buckling shape at $T \gtrsim T_c$ (NT) to a spatially extended wrinkle pattern at the   
far from threshold regime, $T \gg T_c$ (FT), does not stem from non-Hookean response. Instead, the FT wrinkle pattern can be fully characterized by the framework of Hookean elasticity, in which the stress tensor (averaged throughout the thickness of the sheet) has linear dependence on the corresponding strain tensor, but the nonlinear effect of the out-of-plane deflection on the strain within the sheet is taken into consideration. 
This ``mechanically linear'' ({\emph{i.e.}} Hookean stress-strain relationship), yet ``geometrically-nonlinear'' ({\emph{i.e.}} 
rotationally-invariant displacement-strain relationship)
approach to elasticity underlies the celebrated {F\"oppl-von K\'arm\'an} (FvK) equations, and was shown to describe quantitatively fully developed wrinkle patterns in a variety of examples \cite{Toga13,Pineirua13,Taylor15,Taffetani17,Paulsen16,Chopin18,Dinh16,Box19}. 

The essential reason that a Hookean, geometrically-nonlinear framework suffices to explain
the fully developed wrinkle pattern was noted in a seminal 2003 paper of Cerda \& Mahadevan (CM) \cite{Cerda03}. 
Since for very thin sheets the threshold tensile load may be arbitrarily small -- 
more precisely, $T_c \sim Y (\tfrac{t}{W})^2$ \cite{Xin20}, where $Y$ is the stretching modulus --  
the FT regime $T \gg T_c$ is reached 
while the characteristic strain ($T/Y$) remains very small, such that 
Hookean response is a valid approximation 
everywhere in the deformed sheet. Motivated by this observation, 
these authors introduced a model to describe the Hookean FT regime, 
$T_c \ll T \ll Y$, assuming that the formation of wrinkles affects a strong, non-perurbative deviation of the stress field from the planar stress. The mechanism invoked by the CM model     
is strictly distinct from standard ``post-buckling'' analysis, which assumes that the planar stress is only mildly perturbed 
(and therefore characterizes the buckled shape in the NT regime, $T \gtrsim T_c$). 
In the CM model wrinkles are assumed to expand through the whole length $L \gg W$ of the sheet, rather than being confined to the transversely-compressed zones of the planar state, and the wrinkle 
wavelength $\lambda$ and amplitude $A$ 
are determined 
by effective rules that interweave mechanics and geometry, yielding:   
\begin{subequations}
\label{eq:CM-1}
\begin{gather}
\lambda \sim (B/\Keff)^{1/4} = (\frac{Y}{T})^{1/4} L^{1/2} t^{1/2}  \label{eq:CM-1a} \\
A \sim  \sqrt{\nu \frac{T}{Y}} \ \lambda 
\label{eq:CM-1b}
\end{gather}
\end{subequations}
where $B\sim Yt^2$, $Y$, and $\nu$ are, respectively, the bending and stretching moduli and Poisson ratio of the sheet,  
and $\Keff = T/L^2$ is the stiffness of a tension-induced {\emph{``effective substrate''}} that governs 
the resistance to out-of-plane deflection. Briefly, Eq.~(\ref{eq:CM-1a}) reflects a balance between the energetic costs of bending due to transverse undulations ($\propto B (A/\lambda^2)^2$), and the associated pulling of stretched longitudes of the sheet 
between the clamped edges ($\propto T(A/L)^2$). Equation~(\ref{eq:CM-1b}) follows from a second, ``transverse inextensibility'' assumption: 
{\emph{``As the sheet wrinkles in the $y$ direction under the action of a small compressive stress, 
it satisfies the condition of inextensibility ..``}} \cite{Cerda03}. According to this assumption, wrinkles 
{{do not}} emerge to relax (transverse) compressive stress but rather to prevent
transverse strain, $\epsilon_{yy}\approx -\nu T/Y$ (which is the   
transverse contraction of the sheet in the planar, unwrinkled state of the stretched sheet). 

. 






The CM model~\cite{Cerda03} attracted a remarkable level of interest and provoked research activity 
that far exceeded its original realm of application.   
Specifically, the proposal that wrinkle patterns in thin solid bodies should be considered far-from-threshold phenomena and correspondingly be analyzed through a theoretical framework that is sharply distinct from traditional post-buckling methods  
inspired a multitude of experimental and theoretical studies 
in ultarthin sheets subjected to confinement by capillary effects or other forces 
\cite{Vella19,Paulsen19,Bella14,Paulsen20a,Paulsen20b,Vella15,Azadi14,Hure12,Okeily20,Chopin15,Vella18a,Vella18b,Davidovitch19,Tovkach20}.  
In particular, these studies provided strong support to 
the reasoning underlying CM principle~(\ref{eq:CM-1a}) that determines the wrinkle wavelength: a balance between the bending modulus and the stiffness of an effective substrate, 
which may be an actual foundation, 
or induced by a boundary load or curved topography that imply tension perpendicularly 
to the compressed axis \cite{Bico04,Paulsen16}.       

Nevertheless, the validity of the second CM principle~(\ref{eq:CM-1b}) has been challenged by observations that the wrinkle amplitude in experiments and simulations is substantially smaller than this prediction \cite{Nayya11,Healy13,Healy16} (even at the Hookean regime, where the amplitude is observed to increase with applied tension \cite{Pontanescu19}). Furthermore, the mere rationale
of the transverse inextensibility assumption underlying Eq.~(\ref{eq:CM-1b}) is confounding.     
%
According to this assumption, wrinkles emerge to prevent the transverse contraction 
in the bulk of the stretched sheet ({\emph{i.e.}} away from the clumped edges)     
and one would thus expect to observe wrinkling 
even if the pulled edges were not clamped, in which case the whole sheet contracts transversely. Putting it in more formal terms, 
according to Hookean mechanics 
a vanishing transverse strain 
in a sheet under longitudinal tension ($\sxx \!= \!T$)
implies transverse tensile stress ($\varepsilon_{yy} \!=\! 0 \!\Longrightarrow\! \syy\! =\! \nu T$), 
whereas a vanishing transverse compression 
implies a transverse contractive strain ($\syy \!=\! 0 \!\Longrightarrow \!\varepsilon_{yy} \!=\! -\nu T/Y$).       
%
Hence, the CM assumption of vanishing transverse strain 
appears to be at odds with the Poisson effect, which posits that the minimization of elastic energy is attained by eliminating transverse stress. 
Thus, paradoxically, for a sheet under longitudinal tensile load $T \gg T_c$, the CM 
Eq.~(\ref{eq:CM-1b}) implies that the elastic energy of a wrinkled state is larger than the corresponding energy of a 
planar state!

Seeking to clarify these obscure aspects of the CM model,
we revisit in this paper the Hookean FT regime, $T_c \ll T \ll Y$ of this problem. We implement a theoretical 
framework, known as ``FT analysis'' \cite{Davidovitch11,Davidovitch19}, that has been applied successfully for studying various wrinkling problems 
-- a systematic expansion of the FvK energy around the singular limit of a hypothetic, infinitely bendable sheet, which cannot accommodate any compressive stress, and its stress field is the subject of tension field theory (TFT) \cite{Wagner29,Stein61,
MansfieldBook,Pipkin86,Steigman90}. A central part of this approach 
is that the transverse (compressive) stress, rather than the transverse strain, vanishes with the bendability of the sheet, yielding a ``slaving condition'' between the wrinkle amplitude and wavelength. In contrast to previous studies, where FT analysis have been used mostly for highly symmetric systems, amenable to analytic solution of the TFT equations, the current problem does not yield itself to analytic solution, hence we employ the numerical software 
Surface Evolver for finding 
the energetic minimum in the FT regime, where sheets are populated by fine, fully developed wrinkles. Combining theoretical considerations
and numerical simulations we offer a modified version of the CM model for the Hookean FT regime in this problem, which is compatible with the 
Poisson effect's rationale, and revise accordingly its central prediction, Eq.~(\ref{eq:CM-1}).

In Sec.~\ref{sec:th-1} we describe the general principles of TFT and the corresponding FT
analysis of the wrinkle pattern, and provide a revised version of the amplitude-wavelength ratio (Eq.~\ref{eq:CM-1b}) in terms of a ``confinement function'', $\PhiTFT^2 (x)$, that emanates from TFT and characterizes the fraction of transverse arclength that must be ``wasted'' by wrinkles in order to ensure an asymptotically compression-free stress field in the stretched sheet.   
In Sec.~\ref{sec:ft-stress} we present results of our numerical simulations in the FT regime, showing that the emergence of wrinkles comes in tandem with an intricate 
collapse of the transverse compressive stress, whereby the compression level vanishes asymptotically (as $T/T_c \to \infty$ while $T \ll Y$) in comparison to the corresponding planar state, but the spatial extent of the transversely compressed zones is increased. These numerical results substantiate the rationale underlying the FT analysis and highlight similarities and differences with other tensional wrinkling phenomena. In Sec.~\ref{sec:tft} we describe numerical simulations of a hypothetic sheet with finite stretching modulus ($Y \gg T$) but no bending modulus ($B=0$), which allows us to obtain numerically the tension field limit of a compression-free stress field. 
We extract from these simulations the confinement function $\PhiTFT^2(x)$, 
and show how it encapsulates the intrinsic geometrical nonlinearity that stems from infinitesimal out-of-plane deflections on the in-plane transverse strain, even though the exerted longitudinal tensile strain $T/Y$ may be arbitrarily small (so that Hookean mechanics is valid). 
In Sec.~\ref{sec:ft-wavelength} we turn to discuss the 
various aspects of the wrinkle pattern, specifically the wavelength $\lambda$, and the amplitude-wavelength ratio. 
We elucidate some subtlety in evaluating the dependence of the effective, tension-induced stiffness, on the width $W$ and length $L$ of the sheet. While our numerical simulations strongly support the dependence of $\lambda$ on the tensile load $T$ and bending modulus $B$ of the sheet, we argue that the length's dependence predicted in the CM model, Eq.~(\ref{eq:CM-1a}), may not necessarily be valid for $L \gg W $. 
In Sec.~\ref{sec:discussion} we conclude with a summary of results and a discussion of open questions.  



\section{Elements of far-from-threshold analysis} \label{sec:th-1} 
\subsubsection{Overview} \label{subsec:overview}

The various parameters and variables of the model system, as well as the linear (Hookean) relationship between the stress and strain tensors, $\sigma_{ij}$ and $\varepsilon_{ij}$, respectively, 
and the FvK equations of mechanical equilibrium were given in Sec.~II of our preceding paper \cite{Xin20}, where we addressed the planar state 
and its 
buckling instability. Here we follow the same conventions, shown in the schematic Fig.~\ref{fig:schematic}. Specifically, we denote $\tilde{(\cdot)}$ a dimensionless version of a physical parameter or variable $(\cdot)$, where stresses (integrated over the thickness $t$ of the sheet) are normalized by the stretching modulus $Y$, and lengths are normalized by the width $W$. The problem is to find the displacement field, ${\bf u} = (\rmu_x,\rmu_y,\zeta)$ that minimizes the enthalpy 
\begin{equation}
U = E - {\rm Work} 
\label{eq:enthalpy-0}
\end{equation} where the elastic energy $E$ and ${\rm Work}$ are given by 
\begin{gather}
E = \frac{1}{2} \int dxdy  \ B (\nabla^2\zeta)^2  + \sigma_{ij} \varepsilon_{ij} \nonumber \\
{\rm Work} = 2 \cdot T \cdot W  \cdot  \rmu_x(x=L/2,y)  \label{eq:enthalpy-1} \ .  
\end{gather}
Note that since we consider 
small-strain conditions ($\tT \ll 1$), we could simplify the above equations in two ways: first,  mechanically -- by assuming a Hooeakn stress-strain relation (Eq. 2 of \cite{Xin20}), and second, geometrically -- by assuming a small-slope deflection from the plane ($|\nabla \zeta| \ll 1$) and correspondingly using Mong\'e representation in the strain-displacement relation (Eq. 1 of \cite{Xin20}), and approximating the mean curvature by $\tfrac{1}{2}\nabla^2 \zeta$. We also took advantage of the symmetry $x \leftrightarrow -x$. 
In this FvK framework, the nonlinear response emanates solely from the geometrically-nonlinear coupling of out-of-plane displacement to the strain tensor in the sheet, of which the most important component for our problem is: 
\begin{equation}
\varepsilon_{yy} = \partial_y \rmu_y + \tfrac{1}{2} (\partial_y\zeta)^2  \ . 
\label{eq:trans-strain}
\end{equation}
This relation shows that even for large in-plane transverse displacement,  
it is possible for the corresponding strain to be arbitrarily small 
by tuning suitably the deflection from the plane, namely, 
$(\partial_y\zeta)^2 \!\approx\! -2 \partial_y \rmu_y   \ \Rightarrow \  |\varepsilon_{yy}| \!\ll \!|\partial_y \rmu_y|$.   



Using our normalization convention, one readily  
finds that the physics is governed by 3 dimensionless groups:
\begin{equation}
\tT = \frac{T}{Y} \ \ ;  \ \ \epsilon = \frac{B}{TW^2} \ \ ; \ \  \tL = \frac{L}{W} \ . 
\label{eq:non-dim-1}
\end{equation}
The parameter $\tT$ is the characteristic tensile strain imposed on the sheet in the longitudinal axis $\hat{x}$; 
the parameter $\epsilon$ is recognized as the inverse of the ``bendability"\cite{Davidovitch11}, and the parameter $\tL$ is the aspect ratio. We focus on 
the ``corner'' in parameter space $(\tT \!\ll\! 1,\epsilon\! \ll\! 1, \tL \!\gg \!1)$, namely -- the Hookean, yet geometrically-nonlinear response of long, highly bendable ribbons. 

In the preceding paper \cite{Xin20} we showed that the planar state ({\emph{i.e.}} $\zeta=0$) becomes unstable and develop a buckling pattern 
(with a wavelength $\lambda_c \approx W/3$) 
when the exerted tension exceeds a threshold value  
$T_c \sim Y (t/W)^2$. 
Notably, when expressing the system through the dimensionless groups (\ref{eq:non-dim-1}), the threshold occurs along a``vertical'' line ($\epsilon_c,\tT$) in the parameter plane ($\epsilon,\tT)$, where:   
\begin{equation}
\epsilon_c \approx 10^{-6} \ . 
\label{eq:Tc}
\end{equation}
for any $\tL$ larger than about 4. 
Hence, for the rest of this paper, we will refer to the threshold through the value $\epsilon_c$ of the dimensionless parameter $\epsilon$. 
(A reader who finds it more convenient to associate a threshold with the value of the tensile load, may readily convert: $T_c \approx \epsilon_c \cdot B/W^2$).  

Underlying the NT analysis, which is valid for $\epsilon \lesssim \epsilon_c$,
there is an expansion: 
\begin{gather}
U(\tT,\epsilon) = U_{\rm plane}(\tT) + \Delta U 
\label{eq:heirarchy-NT-1}
\end{gather}
where $U_{\rm plane}(\tT)$ is the enthalpy of the planar state, which does not depend on the bending modulus (hence is $\epsilon$-independent), and $\Delta U$ is negative for $\epsilon <\epsilon_c$
such that $|\Delta U| \sim (\epsilon - \epsilon_c)^2 
\ll U_{\rm plane}(\tT)$ for $\epsilon \lesssim \epsilon_c$. 
The buckling shape 
can be found by minimizing $\Delta U$, assuming 
a perturbation with infinitesimal amplitude and 
negligible correction to the planar stress. 

The basic premise of the FT framework is a description of the 
deformed sheet for a regime in the parameter space ($\epsilon^{-1}, \tT$) far beyond the threshold line, {\emph{i.e.}} $\epsilon \ll \epsilon_c$.  
This is done through an 
expansion of the elastic energy 
around the {\emph{singular}} limit $\epsilon \to 0$ for a fixed geometry ($\tL$) and tensile load per thickness ($\tT$). For an experimenter 
whose set-up comprises a single sheet, {\emph{i.e.}} fixed thickness and aspect ratio $\tL$, on which the exerted tensile load is gradually raised or lowered, thereby changing smoothly both $\tT$ and $\epsilon$, such an approach may sound as an obscure mathematical trickery. Nevertheless, we shall show that this theoretical framework bears invaluable advantages for actual computations as well as for conceptual understanding. \\

Underlying the FT analysis (for a sheet with a given $\tL$) there is an expansion: 
\begin{gather}
U(\tT,\epsilon) = \UTFT(\tT) + \Usub(\tT,\epsilon) \nonumber \\
\ \ {\rm s.t.} \ \ \Usub/\UTFT \to 0 \ {\rm as} \ \epsilon \to 0  \ , 
\label{eq:heirarchy-FT-1}
\end{gather}
where the ``dominant'' contribution $\UTFT(\tT)$ is obtained by solving {\emph{tension field theory}} for a hypothetical sheet with finite stretching modulus and zero bending modulus, and $\Usub(\tT,\epsilon) >0$ is a subdominant contribution to the energy, associated with the direct energetic cost of wrinkling: bending the film and deforming the substrate. Crucially, $\UTFT(\tT) < U_{\rm plane}(\tT)$, hence -- for any finite $\tT$ and  sufficiently small $\epsilon$, it is the FT expansion (\ref{eq:heirarchy-FT-1}), rather than its NT counterpart (\ref{eq:heirarchy-NT-1}), which provides a reliable evaluation of the energy, and whose minimization should be used for characterizing the deformation. 

The energetic hierarchy (\ref{eq:heirarchy-FT-1}) entails three principles that comprise the FT expansion: 

\noindent {\emph{(a)}} an asymptotic, compression-free stress field; 

\noindent {\emph{(b)}} ``slaving'' the wrinkle amplitude to its wavelength; 

\noindent {\emph{(c)}} a ``wavelength rule''. 

In the rest of this section we explain these general principles, specializing to our model system. In Secs.~\ref{sec:ft-stress}-\ref{sec:ft-wavelength} we demonstrate through numerical simulations how these elements govern the wrinkle pattern in our problem.





\subsubsection{Asymptotic compression-free stress field}   \label{subsec:TFT} 

In the limit $\epsilon \to 0$, the stress tensor 
in the wrinkled sheet approaches a compression-free limit value. That is, for $0<\epsilon \ll \epsilon_c$ the stress tensor can be approximated as:
\begin{equation}
    \sigma_{ij}(\bfx;\tT, \epsilon) \approx  \sigma_{ij}^{(TFT)} (\bfx;\tT)  \ , 
    \label{eq:stress-TFT-1}
\end{equation} 
where the principal components of the tensor $\sigma_{ij}^{(TFT)} (\bfx;\tT)$ are non-negative everywhere in the sheet. The approximation symbol  
indicates $O(\epsilon^\beta)$ corrections, with $\beta >0$. A central premise of TFT is that the compression-free stress field $\sigma_{ij}^{(TFT)} (\bfx;\tT)$ is a well-defined tensor, 
obtained directly through the energy minimization procedure underlying the 
dominant part $\UTFT(\tT)$ in Eq.~(\ref{eq:heirarchy-FT-1}) -- 
allowing the deformation to have any (wrinkly, highly curved) out-of-plane component while ignoring its energetic cost.  
This amounts to solving the force balance equations for the stress tensor, 
subject to non-negativity of its principal components. 

Being independent on the small parameter $\epsilon$, the tensor $\sigma_{ij}^{(TFT)} (\bfx;\tT)$ 
characterizes the smoothly-varying, {\emph{gross}} features of the wrinkle pattern, the most basic of them is the {\emph{extent}} of the wrinkled zone. The TFT solution marks two regions: 
\begin{gather}
\underline{\rm unwrinkled}: \  \ \  \tfrac{1}{2}L \!- \!x^* \!<\!|x|\! <\!\tfrac{1}{2}L  \nonumber \\ 
\underline{\rm wrinkled}: \ \ \  |x| \!<\! \tfrac{1}{2}L \!-\! x^*  \ . 
   \label{eq:division}
\end{gather}
In the unwrinkled zone, near the clamped edges, $\sigma_{ij}^{(TFT)} (\bfx;\tT)$ is characterized by two   
positive ({\emph{i.e.}} tensile) principal components; 
in the wrinkled, central region, only one principal component is positive and wrinkles undulate along the axis perpendicular to the corresponding principal direction. For set-ups characterized by some spatial ({\emph{e.g.}} axial \cite{Davidovitch11,King12,Vella15,Vella18a} or translational \cite{Chopin15}) symmetry, this direction is 
typically determined by the underlying symmetry, whereas in our problem the clamping of the short edges breaks translational symmetry.  
Nevertheless, we 
expect 
the deviation of the principal directions from $\hat{x},\hat{y}$, correspondingly, to be at most $O(\tT)$, and since we consider only $\tT \ll 1$, we ignore 
such deviations when analyzing 
our numerical simulations. 

Notably, the {\emph{actual}} stress field $\sigma_{ij}(\bfx;\tT, \epsilon)$ is not compression-free, but rather comprises a small residual compressive ({\emph{i.e.}} negative) stress component in the perpendicular axis at the wrinkled zone ($\hat{y}$). Nevertheless, this 
residual 
stress component vanishes as $\epsilon \to 0$. The absence of residual compression from the TFT stress field, $\sigma_{ij}^{(TFT)} (\bfx;\tT)$, 
is intimately related to the fact that 
$\sigma_{ij}^{(TFT)} (\bfx;\tT)$ determines only the gross features of the pattern but 
carries no information on the fine features, specifically the wrinkle wavelength $\lambda$. 
Finally, let us note that the extent of the wrinkled zone, which is determined by $x^*$ in Eq.~(\ref{eq:division}), 
may depend on $\tT$ and $\tL$ (even though the actual dependence turns out to be rather weak), 
but not on $\epsilon$. 
This independence on $\epsilon$ of  
all TFT-derived expressions is crucial for understanding the amplitude-wavelength ``slaving'' condition, which we discuss next.  

\subsubsection{Amplitude-wavelength slaving condition}  \label{subsec:slaving}
Since TFT ignores the energetic cost associated with out-of-plane deflection $\zeta(\bfx)$, any 
contraction of length is facilitated 
by ``wasting'' the excess length through some $\zeta(\bfx)$. Specifically, this means that the transverse strain, $\varepsilon_{yy} (\bfx)$, Eq.~(\ref{eq:trans-strain}), ``decouples'' from 
the corresponding derivative, $\partial_y\rmu_u(\bfx)$, of the transverse displacement (as long as the latter is contractive, {\emph{i.e.}} negative). 
On the other hand, compatibility of the TFT stress field (RHS of Eq.~\ref{eq:stress-TFT-1})
with the limit value of the stress in a Hookean sheet (LHS of Eq.~\ref{eq:stress-TFT-1}) requires that 
$0\approx  \ \syy = Y(\varepsilon_{yy} + \nu \varepsilon_{xx})$ and $T \approx \ \sxx = Y(\varepsilon_{xx} + \nu \varepsilon_{yy})$. 
As a consequence, the TFT solution implies a ``slaving'' condition for
all feasible out-of-plane deflections: 
\begin{equation}
    \varepsilon_{yy} 
    =  \partial_y\rmu_y + \frac{1}{2} (\partial_y\zeta)^2  
   \  \Longrightarrow \  \frac{1}{2} (\partial_y\zeta)^2  =  - \partial_y\rmu_y  -\nu\tT  \ . 
   \label{eq:TFT-00}
\end{equation}
Let us define: 
\begin{gather} 
 \Phi^2(x) \equiv 
\frac{1}{2W}\int_{\frac{W}{2}}^{\frac{W}{2}} (\partial_y\zeta)^2  \ dy
\label{eq:slaving-0} \\
 \tDelta(x) \equiv \frac{1}{W} [\rmu_y(x,-\frac{W}{2}) - \rmu_y(x,\frac{W}{2})] \label{eq:slaving-0a}
\end{gather}
where $\tDelta(x)$ is the contractional  
transverse displacement and $\Phi^2(x)$ is the corresponding ``confinement function'', namely, the excess length wasted by out-of-plane deflections 
(
normalized by the width $W$ of the undeformed sheet). 
From Eq.~(\ref{eq:TFT-00}) we obtain that in the TFT solution, theses 
are related through the relation: 
\begin{gather}
\PhiTFT^2(x) =  
 \tDeltaTFT(x) - \nu \tT  \label{eq:slaving-1}  \ , 
\end{gather} 
Similarly to the convergence of the stress to the compression-free TFT value, Eq.~(\ref{eq:stress-TFT-1}), the functions $\Phi(x)$ and $\tDelta(x)$ of Hookean, bendable sheets converge to their respective TFT values in the limit $\epsilon \to 0$, hence, for $\epsilon \ll 1$ we have that:  
\begin{gather}
    \Phi(x;\tT, \epsilon) \approx \PhiTFT(x;\tT)   \ ; \ 
    \tDelta(x;\tT, \epsilon) \approx \tDeltaTFT(x;\tT)  
    \label{eq:conf-TFT-1}
\end{gather} 
(Note that since $\tDeltaTFT(x)$ and $\PhiTFT(x)$ are determined by TFT, they vanish
at $x \to x^*$, Eq.~(\ref{eq:division})). 
Using a common wrinkling ansatz for the out-of-plane displacement:   
\begin{gather}
\zeta(x,y) \approx A(x)\cdot  g\left(\frac{y}{W}\right)\cdot  \cos\left(\frac{2\pi y}{\lambda(x)}\right) 
\label{eq:ansatz-0} \\
{\rm with:} \ \ g(0)  = 1 \nonumber  \ , 
\end{gather}
where $A(x)$ and $\lambda(x)$ are, respectively, the wrinkle ``amplitude'' and ``wavelength'', and $g(\xi)$ is a slowly-varying ``envelope'' 
(such that ($g'(\xi) \sim O(\epsilon^0)$), Eqs.~(\ref{eq:slaving-0},\ref{eq:conf-TFT-1}) imply a ``slaving'' of the ratio between  
wrinkle amplitude and wavelength of {\emph{actual}} sheets ($0<\epsilon \ll \epsilon_c)$ to the TFT value (of hypothetic sheets with $\epsilon=0$):  
\begin{equation}
\frac{A}{\lambda} \approx C \cdot \PhiTFT(x) \ , 
    \label{eq:slaving-5}
\end{equation}
(where $C$ is some numerical constant, which does not depend on $\epsilon$ or $\tT$).  

As long as Eq.~(\ref{eq:slaving-0}) is satisfied, one may consider     
$\sigma_{ij}^{(TFT)} (\bfx;\tT)$ as 
the stress field in a hypothetic sheet characterized by finite stretching modulus and Hookean stress-strain relation but zero bending modulus. 
We note by passing that since we consider $\tT \!\!\ll\!\! 1$, 
the integrand in Eq.~(\ref{eq:slaving-0}) is 
$\tfrac{1}{2} (\partial_y \zeta)^2 \approx [\sqrt{1+(\partial_y\zeta)^2} - 1]$, that is the 
portion of the transverse arclength ``wasted'' by out-of-plane undulations.   \\

Equations~(\ref{eq:slaving-0},\ref{eq:slaving-1}) highlight two intimately-related flaws 
in 
the original Cerda-Mahadevan model \cite{Cerda03}: \\

$\bullet$ Firstly, Equation~2 of Ref.~\cite{Cerda03} invokes an equality 
of the excess length wasted by wrinkles 
and 
the transverse 
displacement of the free edges, namely: 
\begin{gather}
{\rm CM \ assumption \ (I):}  \label{eq:slaving-1-CM} \\ 
\int_{\frac{W}{2}}^{\frac{W}{2}} dy [\sqrt{1+(\partial_y\zeta)^2} - 1]   \approx 
\rmu_y(x,\frac{W}{2}) - \rmu_y(x,-\frac{W}{2}) \nonumber \\
\left({\rm or:} \ \ \Phi^2(x)   = \tDelta(x) \right)  \ . \nonumber 
\end{gather}
Contrasting this with Eqs.~(\ref{eq:slaving-1},\ref{eq:slaving-5}), we see that CM assumption 
ignores the transverse strain, $-\nu \tT$, which exists in fact also in the fully-developed, compression-free 
wrinkled state. \\

$\bullet$ Secondly, Cerda-Mahadevan assumed that the transverse displacement of the free edges, 
$\tDelta(x) \cdot W$, 
is identical to its counterpart in the planar state, 
namely, 
\begin{gather}
{\rm CM \ assumption \ (II):}  \label{eq:slaving-2-CM} \\ 
\tDelta(x) \!\sim \!\nu \tT \ . \nonumber 
\end{gather}
However, 
Eq.~(\ref{eq:slaving-1}) shows that in order for wrinkles to exist away from the clamped edges
the in-plane transverse displacement of the free edges, $\tDelta(x)\cdot W$,
must {\emph{exceed}} 
the Poisson value, $\nu \tT W$. 
This 
is crucial for understanding the 
very mechanism by which transverse compressive stress is relieved from the planar state: 
further transverse shrinking of the planar projection of the deformed sheet (in comparison to the planar state) is necessary in order to ``make room'' for wrinkles. \\

%


\subsubsection{Effective substrate and wrinkle wavelength}  \label{subsec:wavelength}
For given geometry and loading ({\emph{i.e.}} given $\tL, \tT$), 
there are infinitely many functions $\zeta(x,y)$ that are compatible with Eq.~(\ref{eq:slaving-0}) and 
are therefore legitimate 
candidates to describe the wrinkle pattern.       
For a given $0<\epsilon\ll \epsilon_c$, this degeneracy is lifted by minimizing the residual, sub-dominant contribution $\Usub$ in Eq.~(\ref{eq:heirarchy-FT-1}), associated with the explicit energy cost of out-of-plane deformations in the functional (\ref{eq:enthalpy-0}), 
subject to the slaving constraint (\ref{eq:slaving-5}). 
%
One sub-dominant contribution 
is the bending energy, 
$\propto  B \kappa_{yy}^2$, where 
$\kappa_{yy} \approx \partial^2_y\zeta$ 
is the curvature due to wrinkly undulations, whereas other contributions 
are often gathered into an ``effective substrate'' term, $\propto \Keff \zeta^2$ 
\cite{Cerda03,Paulsen16}, 
with an effective stiffness: 
\begin{equation}
    \Keff = \Ksub + \Kcurv + \Ktens \ . \label{eq:effective-stiff} 
\end{equation}
The various parts of $\Keff$ correspond to a real substrate attached to the sheet ($\Ksub$, {\emph{e.g.}} a heavy liquid bath \cite{Milner89,Pocivavsek08,Huang10,Pineirua13,Tovkach20} or a compliant solid \cite{Bowden98}), a curvature imposed along the axis perpendicular to wrinkly undulations ($\Kcurv$ \cite{Bico04,Paulsen16}), and also a tensile load exerted  along that axis through the boundaries  ($\Ktens$ \cite{Cerda03}).   

In order to elucidate the simultaneous effect of bending rigidity and effective substrate, it is useful to consider
the ansatz (\ref{eq:ansatz-0}) and and amplitude-wavelength slaving (\ref{eq:slaving-5}).  
One may note that the bending energy becomes $\sim B \PhiTFT^2 /\lambda^2 $, whereas the effective substrate energy is $\sim \Keff \PhiTFT^2 \lambda^2$, favoring, respectively, large and small wavelength. 
Such a constrained minimization of the sub-dominant energy yields the scaling relations: 
\begin{gather}
    \lambda \sim \left( \frac{B}{\Keff}\right)^{1/4}
    \label{eq:lambda-0} 
{\rm and}  \ \     \sigma_{res} \sim - \frac{B}{\lambda^2}
\end{gather}
where the residual compressive stress in the undulatory axis ($\sigma_{res} = \sigma_{yy}$ in our problem)  
is obtained by treating it as 
the Lagrange multiplier associated with the slaving constraint (\ref{eq:slaving-5}). 
As was pointed out by Cerda \& Mahadevan \cite{Cerda03}, $\Ktens$ is the only effective stiffness, among the three terms on the RHS of Eq.~(\ref{eq:effective-stiff}), which is operative in our system, making it a primary example -- along with the axisymmetric Lam\'e set-up \cite{Bernal04,Cerda05,Coman08} -- for ``tensional wrinkling'' phenomena.
Nonetheless, a quantitative evaluation of 
$\Ktens$ and $\lambda$ 
beyond the scaling level~(\ref{eq:CM-1},\ref{eq:lambda-0}) requires some subtle considerations,
on which we elaborate below. \\    


In the CM model, the wrinkle amplitude is assumed to vary smoothly between the two clamped edges (where $A=0$), and 
the tension-induced stiffness is therefore estimated as $\Ktens \sim T/L^2$, corresponding to the resistance of a stretched string to deflection (see Eq.~(\ref{eq:CM-1a}) and the subsequent paragraph). More recently \cite{Paulsen16} 
it was pointed out that a quantitative 
estimate of the tension-induced stiffness must take into consideration the actual gradient of the TFT confinement function $\PhiTFT^2(x)$, 
Eq.~(\ref{eq:slaving-1}), so that a more accurate expression for the tension-induced stiffness is:  
\begin{equation}
\Ktens(x) \approx \tfrac{\sigma^{TFT}_{\parallel}(x)}{\ell_{\parallel}^2(x)} \ \ ; \ \ \ell_{\parallel}(x) \equiv 
\left|\frac{\PhiTFT(x)}{\PhiTFT'(x)}\right| \  ,  
\label{eq:Ktens-loc} 
\end{equation}
where $\sigma^{TFT}_{\parallel}(x)$ is the tensile component of the TFT stress tensor ($\sigma^{TFT}_{\parallel}\! \approx\! T$ in our problem), {\emph{i.e.}} {\emph{along}} the wrinkles.  
A spatially-varying stiffness may give rise to a spatially-varying wavelength $\lambda(x)$ \cite{Paulsen16}, or -- 
if the necessary defects are too costly energetically \cite{Taylor15,Jooyan18,Tovkach20} -- to a spatially-uniform wavelength, which reflects a  {\emph{global}} balance of bending and effective substrate energies:

\begin{gather}
    \lambda \approx C_1 \left(\frac{B}{\bKtens}\right)^{\tfrac{1}{4}} 
    \label{eq:lambda-1} \  ;  \ 
\bKtens  = 
T 
\frac{\int_{-\tfrac{1}{2}L + x^*}^{\tfrac{1}{2}L -x^*} dx \PhiTFT'(x)^2} 
{\int_{-\tfrac{1}{2}L + x^*}^{\tfrac{1}{2}L -x^*} dx \PhiTFT(x)^2 }
\end{gather}
where $x^*$ marks the end of the wrinkled zone, Eq.~(\ref{eq:division}), and  
the numerical prefactor $C_1$ is determined by the envelope function $g(y/W)$ of the wrinkle ansatz (\ref{eq:ansatz-0}). 
%
%

Even if the TFT confinement function ${\PhiTFT(x)}$ is known analytically, evaluation of the integral for $\bKtens$ in (\ref{eq:lambda-1}) is hindered by a logarithmic divergence, since 
$\PhiTFT(x) \sim \sqrt{|x - (\tfrac{L}{2} -x^*)|}$, near the end of the wrinkled zone \cite{Davidovitch11,Bella14}. A similar difficulty is 
in axial geometries \cite{Davidovitch12,Taylor15}, where it was found that regularization 
%
%
gives rise to a smooth wrinkle ``foot'' ({\emph{i.e.}} a ``boundary layer'' around $x=\pm (\tfrac{1}{2} L - x^*)$). 
In Sec.~\ref{sec:ft-wavelength} we will discuss a similar effect found upon applying Eq.~(\ref{eq:lambda-1}) to 
our problem.



.

\section{Asymptotic collapse of transverse compression} \label{sec:ft-stress}
Similarly to Ref.~
\cite{Xin20}, we employ Surface Evolver (SE) for numerical simulations, focusing now on the FT regime, 
namely, 
sheets with finite, small bending rigidity, such that
$0<\!\!\epsilon\!\! \ll \epsilon_c$. A characteristic example of such a fully-developed wrinkle pattern is shown in Fig.~\ref{fig:schematic}b.  
In our simulations we implement an equilateral-triangular mesh of density
($total\ area /  cell\ area$) of $6.95\times 10^{5}$, and use the SE built-in method {\emph{``linear\_elastic"}} for computing 
the in-plane strain energy, and the methods {\emph{``star\_perp\_sq\_mean\_curvature"}} and {\emph{``star\_gauss\_curvature"}} for computing the
bending energy. We consider a sheet with a relatively large length-to-width ratio, 
$\tL= 8$, Poisson ratio $\nu = 0.4$, and thickness $t/W = [5\times 10^{-6}, 4\times 10^{-5}]$, 
and vary the exerted tensile load $\tT$. \\ 

\begin{figure}
\centering
\begin{tikzpicture}
    \node at (0,0) {\includegraphics[width=0.8\linewidth]{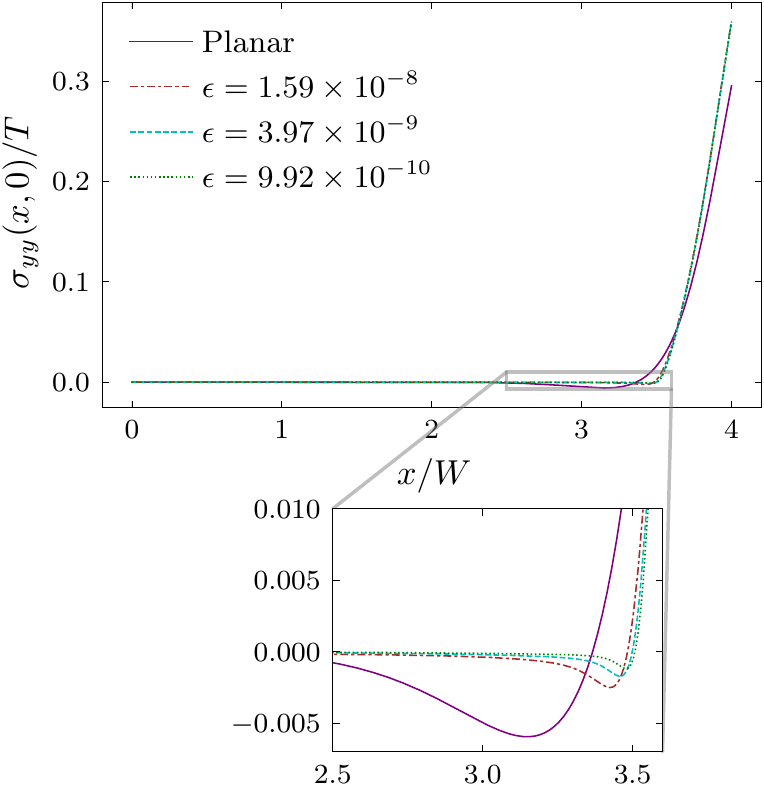}};
    \node at (0, -6) {\includegraphics[width=0.8\linewidth]{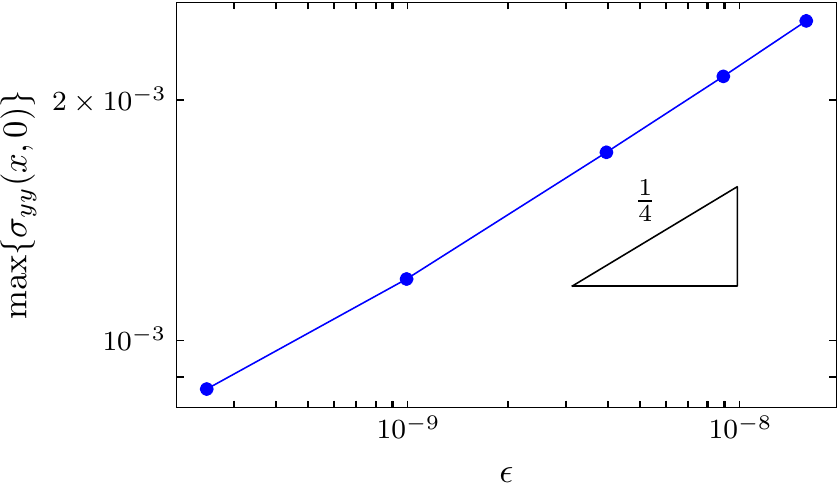}};
    \node at (0, -10.4) {\includegraphics[width=0.8\linewidth]{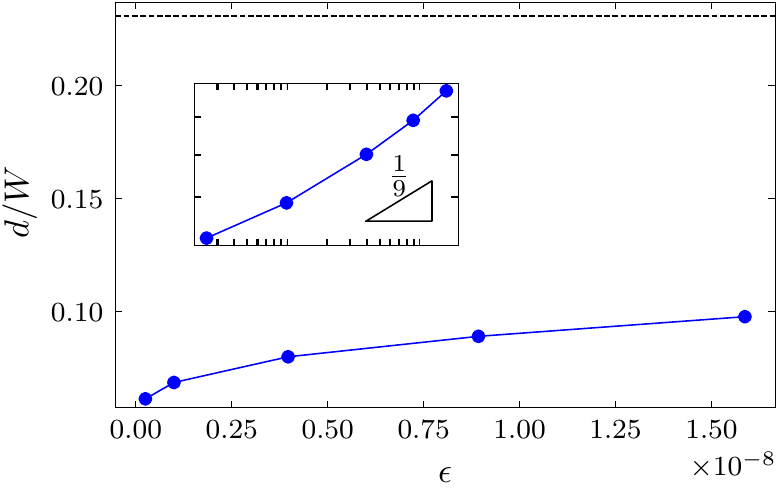}};
    \node at (-3.3, 3.4) {(A)};
    \node at (-3.3, -4.2) {(B)};
    \node at (-3.3, -8.4) {(C)};
\end{tikzpicture}
\caption{
(A) The transverse component of the stress tensor along the midline, $\syy (x,y=0)$ for 
$\tT = 0.01$. 
Shown here are profiles of the planar stress (which is stable when $\epsilon <\epsilon_c$) 
and the stress in a wrinkled state for a few values of the bendability parameters 
($\epsilon/\epsilon_c = 9.3\times 10^{-4}, 0.0037, 0.015$). 
As the bendability increases ($\epsilon \to 0$) we observe a spatially-nonuniform collapse of the compressive stress. 
(B) The dependence of the maximal compression 
($\max_x\{-\langle\syy(x,y)\rangle_y\}$) on $\epsilon$ vanishes at a rate proportional to $\epsilon^{1/4}$. 
(C) The extent of the compressive zone, $d = |x_{max} - x^*|$, 
evaluated as the distance between the points of maximal transverse compression ($x= \tfrac{1}{2}L - x_{max}$)
and zero transverse stress ($x= \tfrac{1}{2}L - x^*$). 
The value of $d$ in the planar state is shown in the dashed horizontal line.} 
\label{fig:stress-collapse}
\end{figure}
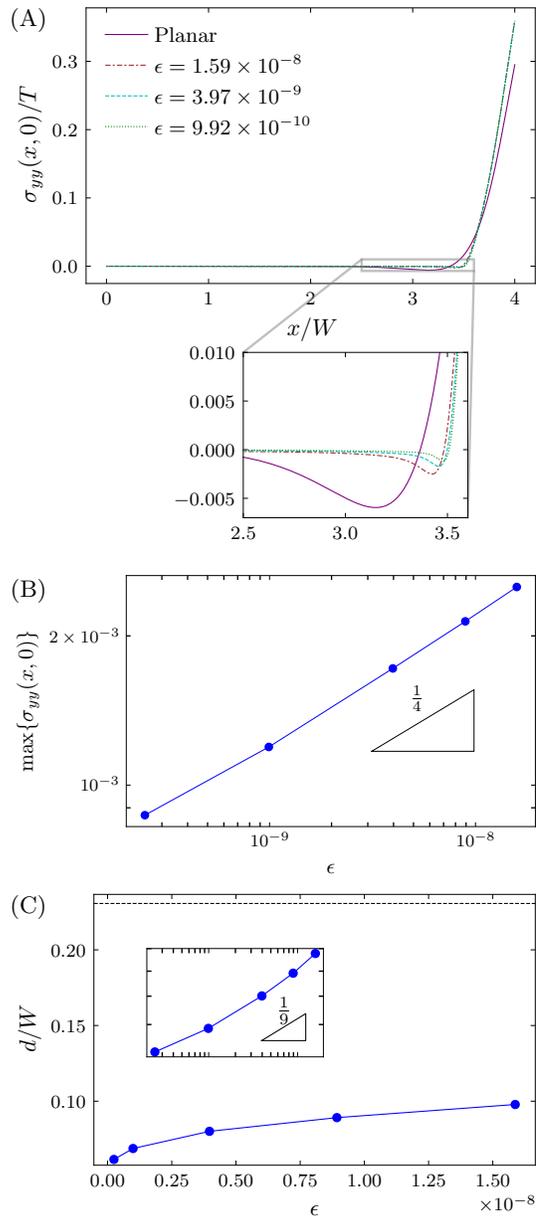

Figure~\ref{fig:stress-collapse}a shows the profile of the transverse stress at the 
midline $\syy(x,y=0)$, for a sequence of values of $\epsilon \approx (9.3\times 10^{-4}, 0.0037, 0.015) \cdot \epsilon_c$. 
The collapse of transverse compression upon decreasing $\epsilon$ is featured by three prominent motifs:

$\bullet$ First, the maximal level of transverse compression, $\max{|\syy(x,y=0)|}$, realized at a distance 
$x_{max}$ from each of the clamped edges, 
decreases substantially from the planar value ($\approx 0.005 \cdot  T$). Furthermore, it vanishes upon decreasing $\epsilon$,  $\max{|\syy|} \sim \epsilon^{1/4}$ (Fig.~\ref{fig:stress-collapse}b). 

$\bullet$ Second, the longitudinal extent of the zone with significant transverse compression also decreases 
substantially in comparison to the planar state. One way to quantify this effect is by considering the distance 
$d = |x_{max} -  x^{*}|$, between the points at which the transverse stress becomes negative and reaches its maximal negative value.  
Figure.~\ref{fig:stress-collapse}c shows that $d$ too vanishes, 
albeit at a much slower rate than $\max{|\syy|}$, namely: $d \sim \epsilon^{1/9}$. 

\begin{figure}
\centering
\includegraphics[width=0.85\linewidth]{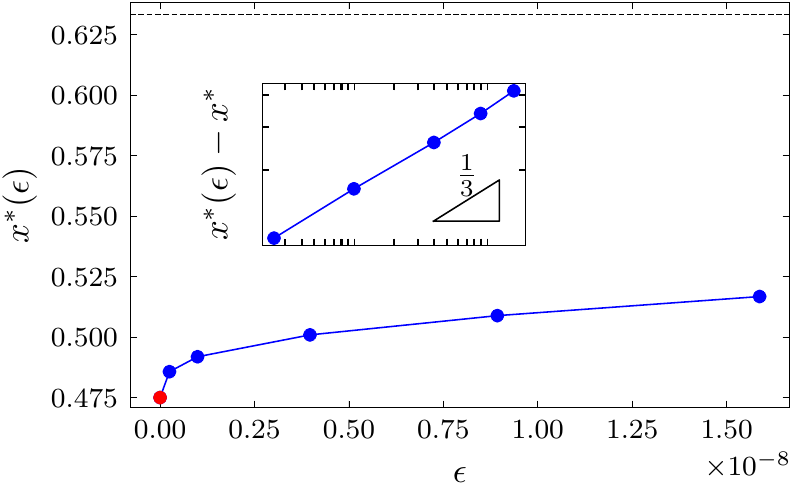}
\caption{ (A) While the compression level in the wrinkled state vanishes asymptotically as $\epsilon \to 0$, 
the transversely-compressed zones gets somewhat closer to the clamped edges. 
We plot here the extent, $x^*(\epsilon)$, of the transversely-tensile zone next to each of the clamped edges ({\emph{i.e.}} the distance from the clamped edge 
at which $\syy(x,y=0)$ 
changes sign for given $\tT = 0.01$). 
The red circle shows the asymptotic value, $x^*$, extracted from the TFT simulations in Sec.~\ref{sec:tft} of a sheet with no bending resistance, and the dashed horizontal line indicates the corresponding length in the planar stress. 
Inset: 
$x^*(\epsilon)$ converges to the TFT value $x^*$  with a residual $\sim \epsilon^{1/3}$.}  
\label{fig:ell-converge}
\end{figure}

$\bullet$ Third, as is shown in Fig.~\ref{fig:ell-converge}, the longitudinal extent $x^{*}$ of the transversely-tensile zones next to the clamped edges, is smaller than its counterpart in the planar state, approaching a finite value as $\epsilon \to 0$. \\

\begin{figure}
\centering
\begin{tikzpicture}
    \node at (0,0) {\includegraphics[width=0.9\linewidth]{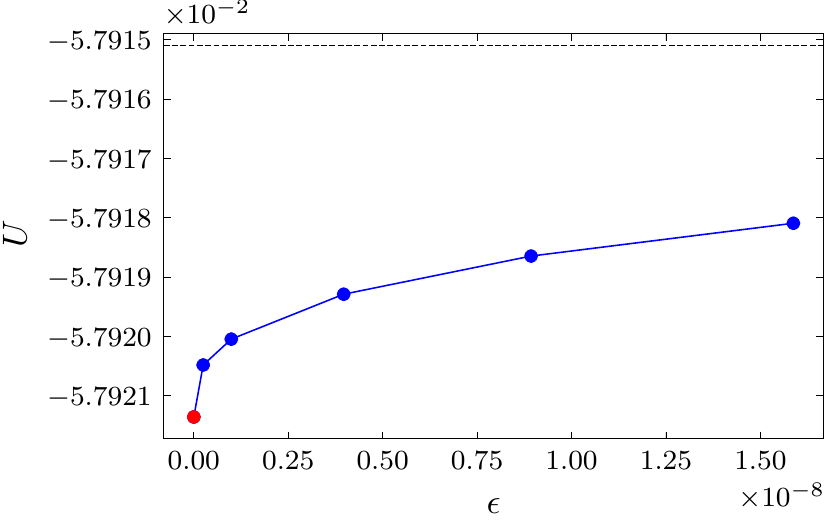}};
    \node at (0, -5) {\includegraphics[width=0.9\linewidth]{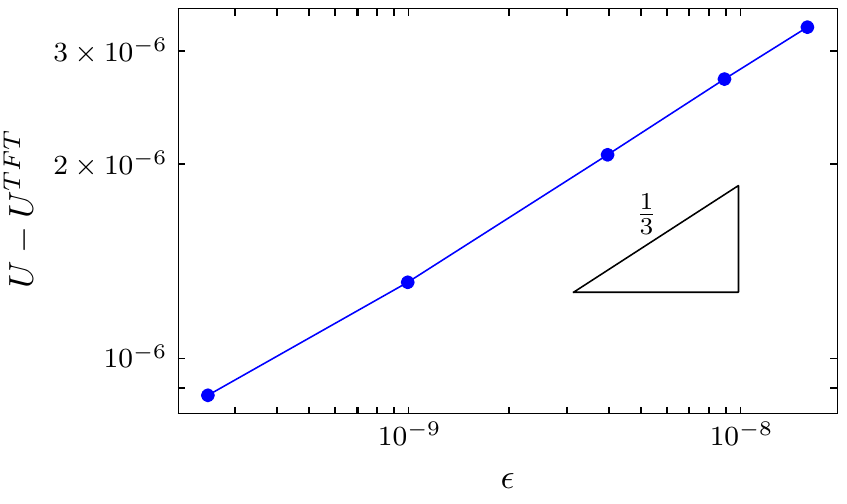}};
    \node at (-3.8, 2.) {(A)};
    \node at (-3.8, -3.) {(B)};
\end{tikzpicture}
\caption{
(A) The energy $U(\epsilon)$ of the wrinkled state for a given tensile load, $\tT=0.01$, and several values of $\epsilon$.  
The red circle indicates the asymptotic energy $U^{TFT}$, extracted from the TFT simulations in Sec.~\ref{sec:tft}, 
and the dashed horizontal line is the energy $U_{plan}$, of the corresponding planar stress. 
(B) the difference $U(\epsilon) - U^{TFT}$, plotted versus $\epsilon$, indicates that the subdominant energy $U_{sub} \sim \epsilon^{1/3}$.
}
\label{fig:energy-converge}
\end{figure}

A central result of our SE simulations is presented in Fig.~\ref{fig:energy-converge}a, where we plot the energy $U(\epsilon)$ (for $\tT = 0.01$) as a function of $\epsilon$. In accord with the scenario described by Eq.~(\ref{eq:heirarchy-FT-1}), this plot shows that the energy is reduced from the value $U_{plane}$ of the planar state (dashed horizontal line), such that the energy gain, $U_{plane}- U(\epsilon)$, 
associated with the formation of a fully-developed wrinkle pattern, approaches a finite value as $\epsilon \to 0$, which we call  
$U_{plane}- U^{TFT}$, attributing it to the prevalence of tension field theory in the high bendability limit, $\epsilon \to 0$. 
Assuming that 
the sub-dominant energy, $U_{sub} =U(\epsilon)-U^{TFT}$, is determined by a work 
of a virtual 
compressive load, whose magnitude is equal to the residual compressive stress 
$\sim \max{|\syy|}$, that exists in a zone of length $d$,   
one may expect that $U_{sub} \sim d \cdot \max{|\syy|} \sim \epsilon^{13/36}$. 
This is rather close to the scaling 
extracted from direct evaluation of the energy, $U_{sub}\sim \epsilon^{1/3}$ (Fig.~\ref{fig:energy-converge}b).

Taken together, these numerical observations reverberate the universal scenario outlined in Sec.~\ref{subsec:overview}
for NT-FT transition between the parameter regime, $\epsilon \lesssim\epsilon_c$, 
which is governed by the transvesely-compressed planar stress, and 
the regime, 
$\epsilon \ll \epsilon_c$,
where the fully-developed wrinkle pattern enables the stress field to approach a distinct, compression-free profile, thereby entailing a finite,  
$\epsilon$-independent energetic gain, $U_{plane}-U^{TFT}$. 





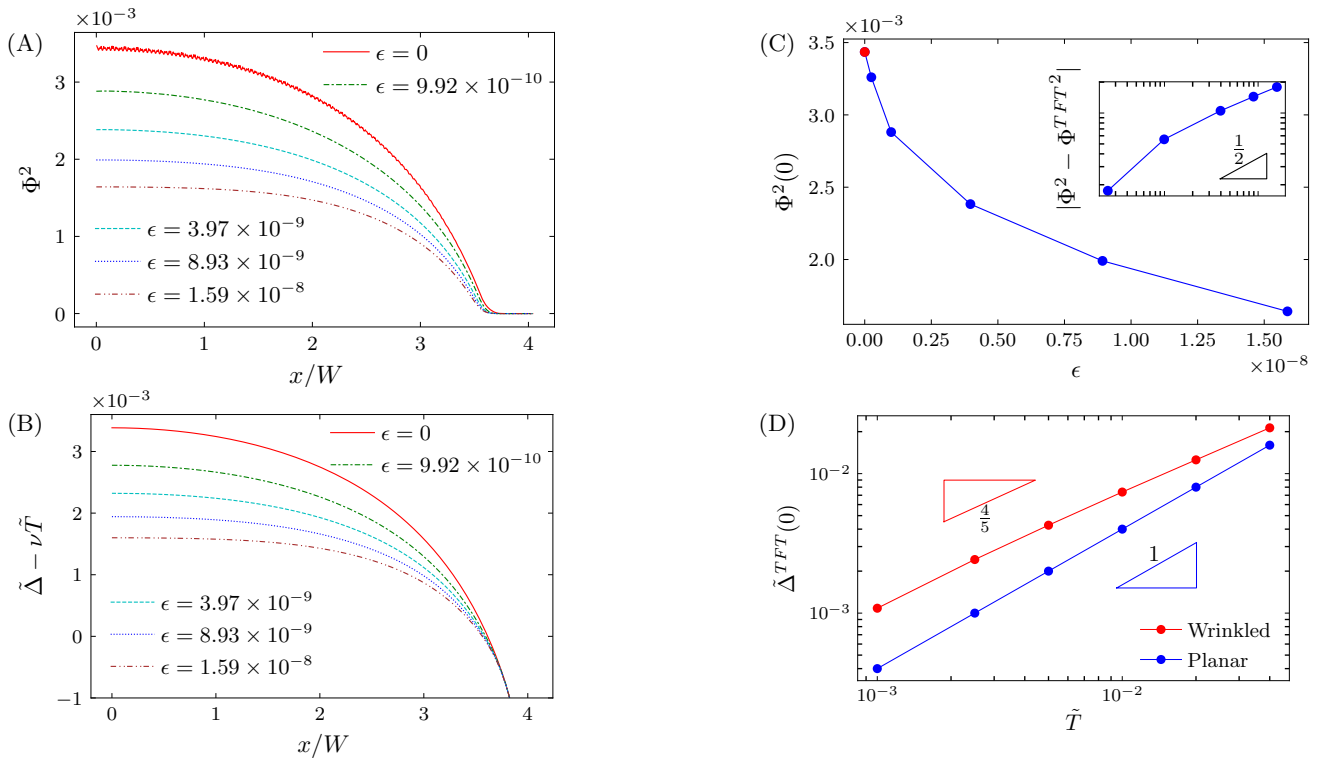
\begin{figure*}
\centering
\begin{tikzpicture}
    \node at (-5,2.5) {\includegraphics[width=0.4\linewidth]{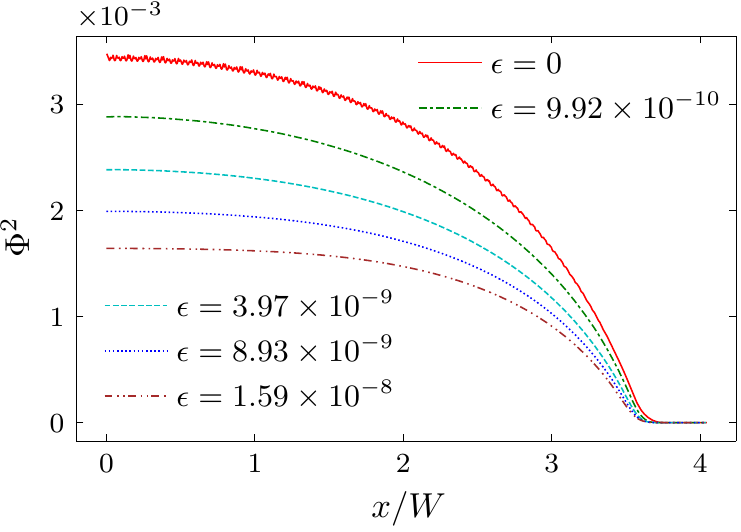}};
    \node at (-5, -2.5) {\includegraphics[width=0.4\linewidth]{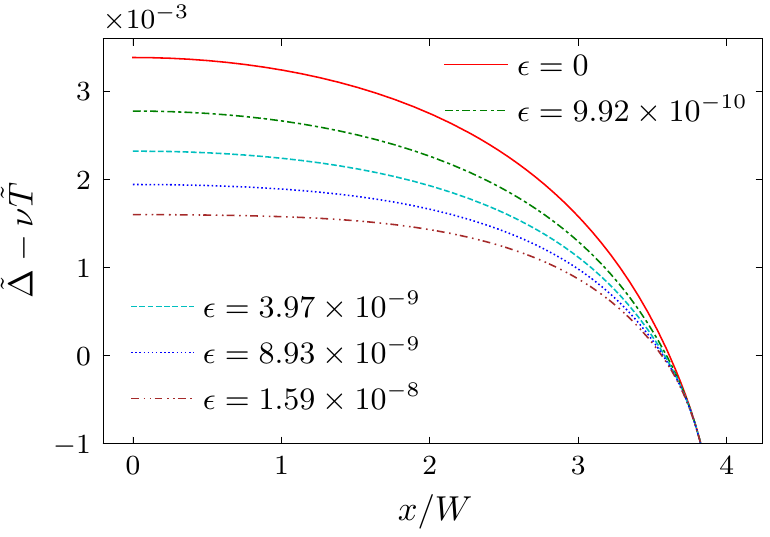}};
    \node at (5, 2.5) {\includegraphics[width=0.4\linewidth]{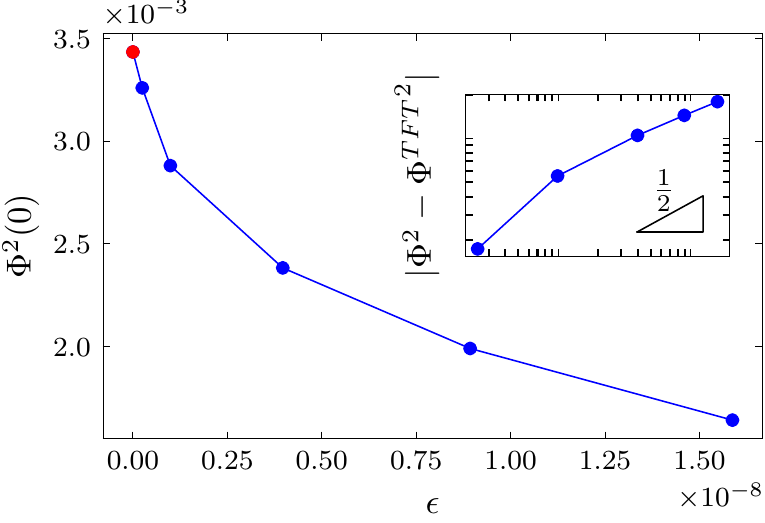}};
    \node at (5, -2.5) {\includegraphics[width=0.4\linewidth]{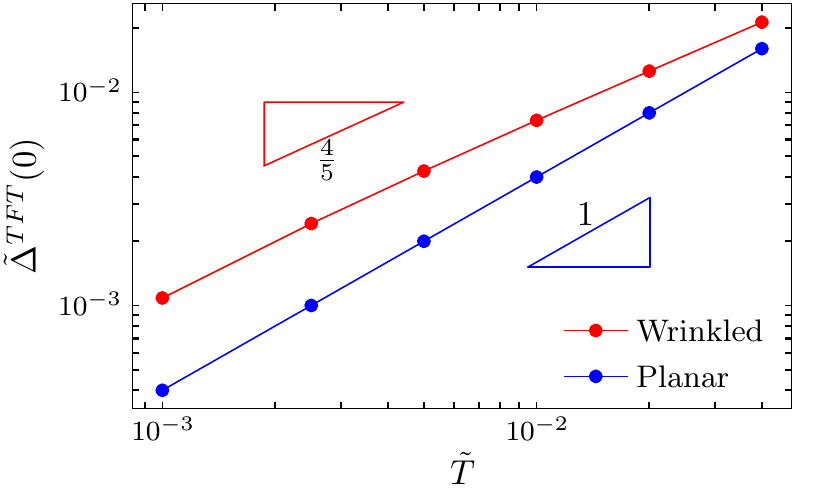}};
    \node at (-8.5, 4.55) {(A)};
    \node at (-8.5, -0.5) {(B)};
    \node at (1.5, 4.55) {(C)};
    \node at (1.5, -0.5) {(D)};
\end{tikzpicture}
%
\caption{
(A) The confinement function $\Phi^2(x)$ for $\tT = 0.01$, extracted by computing the excess length
(RHS of Eq.~(\ref{eq:slaving-0}) in the SE simulations. The red curve is extracted from the TFT solution 
({\emph{i.e.}} a sheet with no bending resistance), 
and the other curves are extracted from the simulations described in Sec.~\ref{sec:ft-stress} for several values of $0\!<\!\epsilon \!< \!\epsilon_c$. 
(B) The transverse contraction of the planar projection, $\tDelta(x)$, extracted by evaluating the RHS of Eq.~(\ref{eq:slaving-1}), 
from the same simulations as in panel A. We subtract $\nu \tT$ from the computed $\tDelta(x)$, in order to allow easy comparison with the computed confinement function $\Phi^2(x)$ in A, and thereby examining the compression-free constraint, Eq.~(\ref{eq:slaving-1}), in the TFT limit (red curves, $\epsilon = 0$), and how this condition is approached as $\epsilon \to 0$.    
(C) The computed value of the confinement function at the center of the sheet, $\Phi^2(x=0)$, plotted {\emph{versus}} $\epsilon$, 
exhibits convergence to the value $\PhiTFT^2(x=0)$ 
of TFT solution (red dot). 
Inset: the convergence to $\PhiTFT^2(x=0)$
is characterized by a residual, $\sim \epsilon^{1/2}$.  
(D) The transverse contraction of the planar projection at the center of the sheet in the TFT solution, $\tDeltaTFT(x=0)$, plotted {\emph{versus}} 
$\tT$, exhibits a nonlinear dependence on $\tT$. 
For reference, we show also the analogous quantity extracted from the planar state \cite{Xin20}, which (for sufficiently large $\tT$) is given by the 
Poisson value $\nu \tT$.} 
\label{fig:confinement-func}
\end{figure*}

\section{Tension field theory -- the limit of compression-free stress field}  \label{sec:tft} 
The observations described in the preceding section provide strong evidence to the prevalence of 
an asymptotic, compression-free stress field, which underlies key features of the fully-developed wrinkle pattern. 
Nonetheless, such a stress field, which is the subject of TFT, can be realized only by a hypothetical sheet with no bending rigidity, 
and hence cannot be attained by simulating a physical Hookean sheet 
({\emph{i.e.}} $\epsilon>0$), no matter how small $\epsilon$ is.  
In this section we seek to resolve this hurdle through SE simulations of precisely such a hypothetical sheet, free of bending rigidity, from which we extract directly the asymptotic stress field and the constraints imposed on the wrinkle pattern.


For a sheet with no bending rigidity, but finite stretching modulus $Y$, 
only tensile stress can be accommodated at mechanical equilibrium. Furthermore, since curvature comes at no energetic cost, 
even an infinitesimal amount of compression is fully relaxed by energy-free, out-of plane undulations. 
The only (non-physical) mechanism limiting the scale of 
such undulations is the mesh size used in the simulation. Hence, as the mesh is made denser, the shape appears to be rougher.
Nevertheless, we show in App.~\ref{sec:App-TFT} that the increasing corrugation does not affect the macro-scale features of the deformation, 
nor does it affect the stress components, all of which appear to converge to well-defined values, independent on the mesh density.  

Our SE simulations of the TFT solution enable us to compute {\emph{directly}} the dominant energy, $U^{TFT}$ in Eq.~(\ref{eq:heirarchy-FT-1}), denoted by red circle in Fig.~\ref{fig:energy-converge}, rather than by 
extrapolating the limit value, $U^{TFT} = \lim U (\epsilon \to 0)$ from results of SE simulation at finite values of $\epsilon$. This is crucial for our ability to compute the scaling, $ U(\epsilon) - U^{TFT} \sim \epsilon^{1/3}$ of the subdominant energy (inset of Fig.~\ref{fig:energy-converge}), which we discussed above, as well as 
the asymptotic extent of the wrinkled zone in the sheet (red circle in Fig.~\ref{fig:ell-converge}) 
and how it is approached as $\epsilon \to 0$ (inset of Fig.~\ref{fig:ell-converge}). 

The most valuable reward for solving numerically the compression-free stress 
is a direct computation of 
the conjugated excess length, namely, the confinement function, $\PhiTFT^2(x)$, as well as the 
transverse contraction of the planar projection, $\tDeltaTFT(x)$, from Eqs.~(\ref{eq:slaving-0}) and (\ref{eq:slaving-0a}), respectively.  
Let us elaborate on several important insights that are 
revealed 
in Fig.~\ref{fig:confinement-func}.  \\  

\begin{figure}
\centering
\begin{tikzpicture}
\node at (-0.01, -0) {\includegraphics[width=0.98\linewidth]{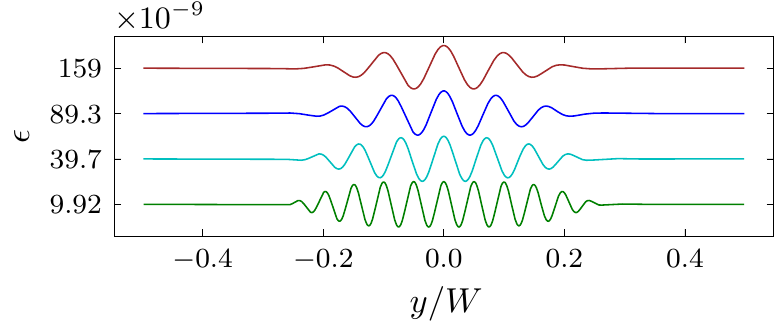}};
\node at (-0, -4.5) {\includegraphics[width=\linewidth]{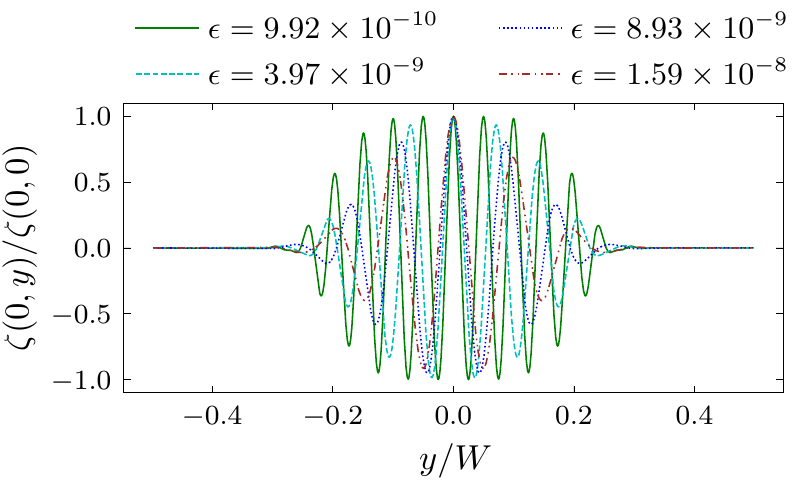}};
\node at (-0.01, -9.4)  {\includegraphics[width=0.98\linewidth]{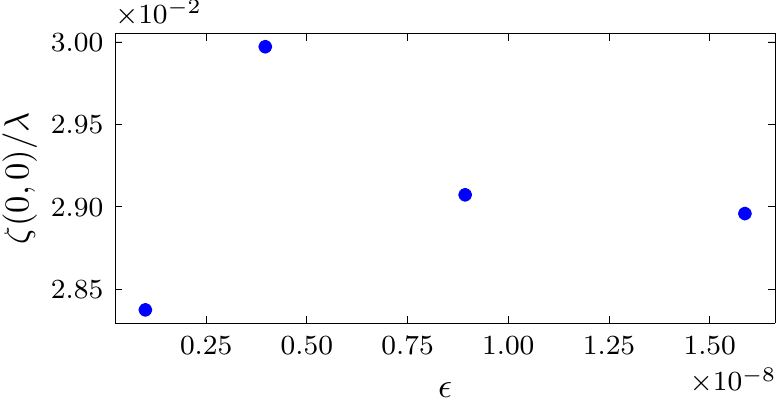}};
\node at (-0.01, -13.8)  {\includegraphics[width=0.98\linewidth]{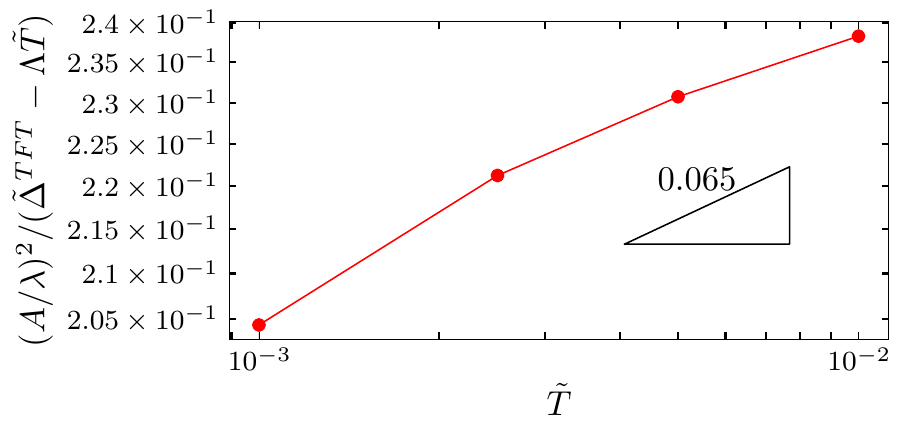}};
\node at (-4, 1.2) {(A)};
\node at (-4, -2.3) {(B)};
\node at (-4, -7.6) {(C)};
\node at (-4, -11.6) {(D)};
 \end{tikzpicture}
\caption{
(A) Transverse profile of the wrinkles, measured at the center of the sheet, $\zeta(x=0,y)$ for $\tT=0.01$ and several values of $\epsilon$. 
The profiles are made 
discernible by shifting them vertically, and multiplying their
amplitude 
by an arbitrary factor. (B) 
Normalizing each profile by the maximal amplitude, $\zeta(0,0)$, the wrinkly profiles appear to be confined to the central half of the sheet width, and be enveloped by a slowly-varying function $g(y/W)$ that becomes constant at $|y| < \tfrac{1}{2}W$ as $\epsilon \to 0$.  
(C) Plotting the amplitude-wavelength ratio (extracted from their respective values at the center) {\emph{versus}} $\epsilon$, for a given value of $\tT = 0.01$, we find that the ratio is not affected by $\epsilon$, in accordance with the FT framework.   
(D) Dividing the amplitude-wavelength ratio by the TFT confinement function, $\PhiTFT^2(x=0)$ (Fig.~\ref{fig:confinement-func}), we find a weak dependence on $\tT$. This discrepancy with the FT framework may be attributed to the modulations of the amplitude. }
\label{fig:profile}
\end{figure}

$\bullet$ The mere existence of a well-defined confinement function $\PhiTFT^2(x)(\tT,\tL)$ (red curve in 
Fig.~\ref{fig:confinement-func}a)
proves the basic premise of the FT framework underlying the CM model (Eq.~\ref{eq:CM-1b}). 
Namely, 
for a given geometry ({\emph{i.e.}} $\tL$), the exerted load $\tT$ determines the transverse arclength wasted by out-of-plane deflections, thereby   
enforcing a finite, $\epsilon$-independent ratio between the asymptotically-vanishing wavelength and amplitude of wrinkles.
The numerically-evaluated confinement function in
Fig.~\ref{fig:confinement-func} is analogous to similar constructs in analytically-tractable models \cite{Davidovitch11,King12,Vella18b,Chopin15}. \\

$\bullet$ As we argued in Sec.\ref{subsec:slaving}, the collapse of transverse (compressive) stress {\emph{does not}} 
require an equality of $\Phi^2(x)$ (\ref{eq:slaving-0}) and $\tDelta(x)$ (\ref{eq:slaving-0a}), which was postulated in the original CM model \cite{Cerda03}, and would have implied a vanishing transverse strain. Instead, collapse of transverse compression requires  
$\Phi^2(x) -\tDelta(x) 
\approx  \nu \tT$, Eqs.~(\ref{eq:slaving-1},\ref{eq:conf-TFT-1}). 
The apparent equality of the red curves in Fig.~\ref{fig:confinement-func}a and Fig.~\ref{fig:confinement-func}b 
supports Eq.~(\ref{eq:slaving-1}), 
thereby proving that underlying the fully-developed wrinkle pattern there is a collapse of transverse compression (stress) rather than vanishing transverse strain.     \\ 

$\bullet$ An important property of TFT, revealed by Figs.~\ref{fig:confinement-func}b and \ref{fig:confinement-func}c, 
is that $\tDeltaTFT(x) \sim \tT^{4/5} > \nu \tT$. 
Namely, the planar projection of the deformed sheet is narrower than its counterpart in the planar state (which is  
in turn directly determined by the Poisson effect). 
Furthermore, the observation that the transverse contraction $\tDeltaTFT(x)$, 
as well as the confinement function $\PhiTFT^2(x)$, are {\emph{not}} proportional to $\tT$, indicates that TFT is a {\emph{nonlinear}} theory of the in-plane strains. 
This is notable, since TFT has a similar formal structure to the planar state solution, 
which is obviously linear in $\tT$. Namely, both theories amount to 
minimizing an energy functional, expressed solely through a quadratic (Hookean) form of in-plane displacement field ($\rmu_x,\rmu_y$), 
where TFT is supplemented by the compression-free constraint (Sec.~\ref{subsec:TFT}). The observation that $\tDeltaTFT(x)$ is not proportional to $\tT$
points to the obscure way by which the compression-free constraint on the TFT stress field (\ref{eq:stress-TFT-1}) 
embodies the geometrical nonlinearity (\ref{eq:trans-strain}), even though the actual out-of-plane displacement $\zeta(x,y)$ is absent from
the TFT calculation.   \\


$\bullet$ The rest of the curves in Figs.~\ref{fig:confinement-func}a,\ref{fig:confinement-func}b ({\emph{i.e.}} other than the red solid) 
show the analogous quantities, extracted from the \mbox{finite-$\epsilon$} SE simulations that were described in the preceding section. In accord with
Figs.~\ref{fig:stress-collapse},\ref{fig:ell-converge},\ref{fig:energy-converge}, which indicated convergence to the TFT limit as $\epsilon \to 0$, 
we observe that the constraints imposed by TFT on the transverse contraction and wasted arclength are reached asymptotically by physical sheets upon increasing their bendability, therey proving Eq.~(\ref{eq:conf-TFT-1}). 
Figure~\ref{fig:confinement-func}d suggests that the 
convergence of these features to the TFT limit values is $\propto \epsilon^{1/2}$, somewhat more rapidly than the 
convergence of the stress, energy, and the longituidinal extent of the wrinkled zone to their respective TFT values.













\section{The wrinkle pattern} \label{sec:ft-wavelength}

Having established the existence of an asymptotic, compression-free stress field, Eq.~(\ref{eq:stress-TFT-1}), 
and the conjugate amplitude-wavelength slaving constraint, Eq.~(\ref{eq:slaving-1}), imposed by TFT, 
we are now at a position to study the fine features of the wrinkle pattern, following the prescription laid out in  
Sec.~\ref{subsec:wavelength}. Aiming to examine the validity of the CM scaling law, Eq.~(\ref{eq:CM-1}), we start by comparing our numerical observation with the ansatz (\ref{eq:ansatz-0}), and then proceed to address the wavelength $\lambda$. 
As we will argue below, our SE simulations enable us to analyze how $\lambda$ varies with bendability and tensile load ({\emph{i.e.}} 
$\epsilon^{-1}$ and $\tT$), but not the manner in which 
the wrinkle pattern varies with $\tL$, 
a task that requires substantial computational power that is beyond the scope of the current paper. A consequence of this shortcoming is 
that we cannot address directly the scaling relation $\lambda \sim L^{1/2}$, predicted in the CM model (\ref{eq:CM-1a}). 
We explain the rationale of this prediction from the perspective of the FT framework, 
and discuss how future simulations of sheets with $\tL \gg 1$ may support or revoke this predicted scaling. 


\subsection{Wrinkling ansatz and the amplitude-wavelength slaving condition} 
Figure~\ref{fig:profile}a shows the transverse profile of the deformed sheet at the center, $\zeta(x=0,y)$, for a given value of $\tT = 0.01$ 
and several values of $\epsilon < \epsilon_c$. 
(Note that amplitudes are not up-to-scale, in order to make the profiles fit into a single figure). Two noteworthy features are: {\emph{(i)}} the characteristic wavelength increases with $\epsilon$; {\emph{(ii)}} the wrinkle amplitude is modulated across the width of the sheet, reaching a maximal value at the centerline ($y=0$). The transverse modulation of the amplitude is further highlighted in Fig.~\ref{fig:profile}b, where we re-plot the wrinkle profiles, normalizing each of them by its maximal amplitude, $\zeta(x=0,y=0)$.


The numerical finding shown in Fig.~\ref{fig:profile}b supports the wrinkle ansatz (\ref{eq:ansatz-0}), suggesting that:
{\emph{(i)}} the transverly-confined zone in the sheet does not extend throughout the whole width, but is instead limited to the central half of the width. {\emph{(ii)}} the transverse undulations of the wrinkle amplitude reflect a slow convergence to an $\epsilon$-independent envelope. More precisely: 
\begin{gather}
\zeta(x,y) \approx \PhiTFT(x)\cdot g(y) \cos(\frac{2\pi y}{\lambda})\cdot (1 + O(\epsilon^{\beta}))   \nonumber \\
{\rm where:} \ \ g(y) \approx \Theta(|y|-\frac{W}{2})  .  
\end{gather} 
Here, $\Theta (y)$ is the Heavyside function, and $\beta$ is some positive constant, whose actual value is beyond the scope of this paper. 

In Fig.~\ref{fig:profile}c we plot the amplitude-wavelength ratio (which we determine for each profile through the largest amplitude, $\zeta(0,0)$), for the various profiles in Fig.~\ref{fig:profile}a). In accordance with the basic paradigm of the FT framework (Sec.~\ref{subsec:slaving}),
we find that 
this ratio is essentially independent on the bending modulus of the sheet ({\emph{i.e.}} $\epsilon$). 
Furthermore, dividing the amplitude-wavelength ratio at a given $\tT$ by $\PhiTFT^2(x=0,\tT)$, and plotting the result 
{\emph{versus}} $\tT$ (Fig.~\ref{fig:profile}d), we find a good agreement with the amplitude-wavelength slaving condition (\ref{eq:slaving-5}) that we obtained in Sec.~\ref{subsec:slaving}.  
(The slight deviation from constancy, $\sim \tT^{0.06}$, may be attributed to the modulation of the amplitude across the width, and to the fact that we determine the amplitude-ratio only through the central wrinkle).   
Notably, the nonlinear dependence of $(A/\lambda)^2$ on $\tT$ even though $\tT \ll 1$ and the simulated sheets are Hookean 
is in clear contradiction to Eq.~(\ref{eq:CM-1b}) of the CM model. 
Thus, Fig.~\ref{fig:profile}d highlights the two intimately-related drawbacks in Eq.~(\ref{eq:CM-1b}), 
which we mentioned already in our discussion in Sec.~\ref{subsec:slaving}: 

{\emph{(i)}} The amplitude-wavelength ratio is determined by the collapse of
transverse compressive stress, hence by the TFT solution, 
and not by a vanishing transverse strain. 

{\emph{(ii)}} The amplitude-wavelength ratio is nonlinear function of the exerted strain $\tT$ 
even for $\tT \ll 1$, 
thereby reflecting 
the geometrically nonlinear nature of TFT.




\begin{figure}
\centering
\begin{tikzpicture}
    \node at (0,0) {\includegraphics[width=0.8\linewidth]{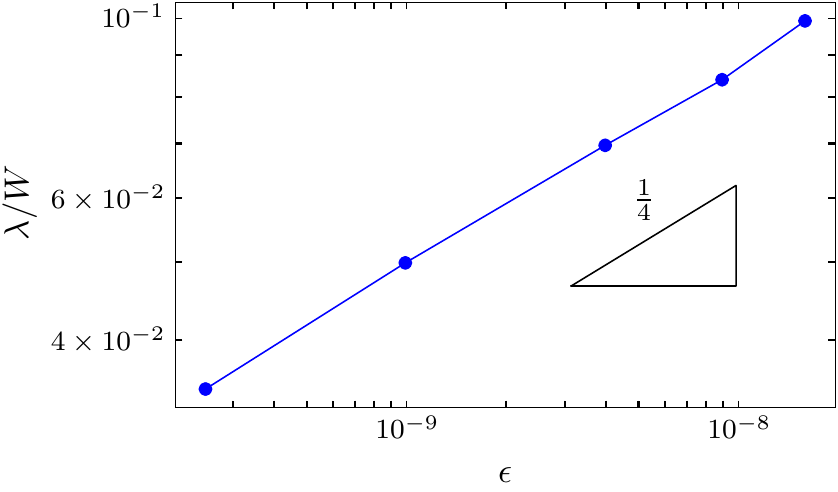}};
    \node at (0,-4) {\includegraphics[width=0.8\linewidth]{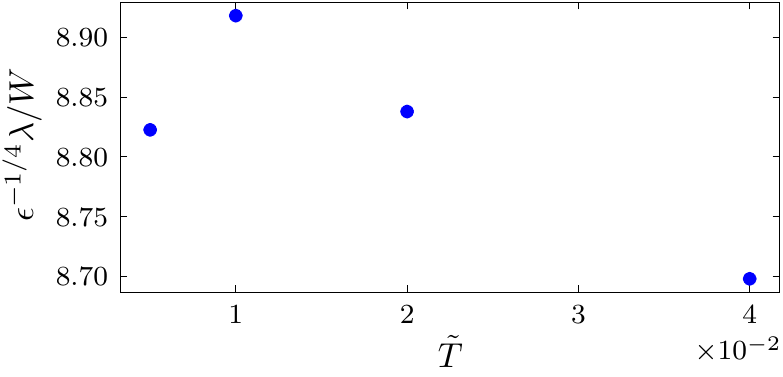}};
    \node at (-3.3, 1.85) {(A)};
    \node at (-3.3, -2.5) {(B)};
\end{tikzpicture}
\caption{
(A) The wavelength $\lambda$ (measured at the center of the sheet) for a fixed value of $\tT = 0.01$ and a few values of $\epsilon$ (log-log plot) exhibits the scaling $\lambda \sim \epsilon^{1/4}$ predicted by the CM model (Eq.~\ref{eq:CM-1a-dimensionless}). 
(B) Plotting $\epsilon^{-1/4}\lambda$ {\emph{versus}} $\tT$, we find a nearly constant value, 
indicating that the exerted tensile strain $\tT$ 
does not affect the wavelength.}
\label{fig:wavelength}
\end{figure}




\subsection{How do bending rigidity and tension 
affect the wavelength ?} 
In order to analyze the wrinkle wavelength $\lambda$, it is useful to express the prediction (\ref{eq:CM-1a}) of the CM model using the three dimensionless groups, $\epsilon,\tT$, and $\tL$: 
\begin{equation}
{\rm CM \ prediction \ (dimensionless):}  \ \ 
\frac{\lambda}{W} \sim \epsilon^{1/4} \cdot \tL^{1/2}     \ . 
\label{eq:CM-1a-dimensionless}
\end{equation}
Notably, the CM model predicts that $\lambda$ depends on the ratio between the bending modulus and exerted tensile load (through $\epsilon$), 
as well as the rectangular shape (through $\tL$), but is indifferent to the exerted tensile strain $\tT$. Recalling our discussion in Sec.~\ref{subsec:wavelength}, we note that the dependence of $\lambda$ on $\epsilon$ follows directly from Eq.~(\ref{eq:lambda-1}), since the integral 
expression for $\Ktens$ is fully determined by the TFT confinement function, $\PhiTFT^2(x)$, 
which -- being a product of TFT -- can depend only on $\tT$ and $\tL$. 
However, the dependence of $\lambda$ on $\tT$ and $\tL$ may be more complicated, since -- as we have seen already in analyzing the amplitude-wavelength ratio -- the geometrical nonlinearity underlying TFT may impart a nonlinear dependence of $\PhiTFT(x)$ on these parameters. 
Being limited to a single value of $\tL =8$, our SE simulations enable us to address the dependencies of $\lambda$ on $\epsilon$ and $\tT$, but not on $\tL$ (on which we will comment in the following subsection).

Figure~\ref{fig:wavelength} shows the wavelength $\lambda$, extracted from our SE simulations. For consistency, we determine $\lambda$ in each wrinkled sheet as $|y^+ - y^-|$, where $y^+, y^-$ are the closest points to the center at which the deflection vanishes, {\emph{i.e.}} $\zeta (x=0,y^{\pm}) =0$. In Fig.~\ref{fig:wavelength}a we focus on a single value of $\tT=0.01 $ and plot $\lambda$ {\emph{versus}} $\epsilon$, 
finding an excellent agreement with the CM prediction~(\ref{eq:CM-1a-dimensionless}). In Fig~\ref{fig:wavelength}b we plot 
$\lambda \cdot \epsilon^{-1/4}$ {\emph{versus}} $\tT$, and find no apparent dependence on $\tT$, again in excellent agreement with the prediction of the CM model. This finding indicates that although the magnitude of the TFT confinement function $\PhiTFT^2(x)$ is a nonlinear function of the exerted tensile strain, $\tT$, its 
spatial variation along the sheet 
is barely affected by $\tT$.


Attempting to obtain a quantitative test for the prediction~(\ref{eq:lambda-1}), one may naturally seek to 
employ the confinement function $\PhiTFT^2(x)$ found in our numerical solution of the TFT in Sec.~\ref{sec:tft} (Fig.~\ref{fig:confinement-func}d,d) for several values of $\tT$, 
and evaluate the corresponding integrals that define $\Ktens$. 
However, 
as we indicated in Sec.~\ref{subsec:wavelength} this scheme is readily stymied due to the logarithmic divergence of the integral in the numerator of $\Ktens$. (Note that, as is evident in Fig.~\ref{fig:confinement-func}a, $\PhiTFT^2(x) \propto x-(\pm\tfrac{1}{2}L-x^*)$, at the vicinity of the
boundary of the transversely-confined zone, yielding $\PhiTFT'(x)^2 \propto [x-(\pm\tfrac{1}{2}L-x^*)]^{-1}$). This divergence indicates that another 
physical effect, which is not accounted for in the balance of bending and stretching energies underlying Eq.~(\ref{eq:lambda-1}), becomes significant at 
$|x| \lesssim \tfrac{1}{2}L -x^*$. A similar phenomenon has been found in tensional wrinkling of an annular sheet (the Lam\'e problem) \cite{Davidovitch11}, where numerical simulations showed that divergence is inhibited through the formation of a partially-compressed boundary layer (whose width decreases slowly 
with $\epsilon$) 
\cite{Taylor15},
although another regularization mechanism that involves wrinkle cascades has also been proposed \cite{Bella14}. While a 
suitably regularized calculation of the integral in Eq.~(\ref{eq:lambda-1}) is beyond the scope of our paper, 
we note that the solution of the Lam\'e set-up suggests that sufficiently far from threshold 
the wavelength retains the scaling $\lambda \sim \epsilon^{1/4}$ as if the integral in Eq.~(\ref{eq:lambda-1})) was convergent (albeit with a numerical 
prefactor whose evaluation requires regularization). Consequently, since our numerical results support the 
scaling $\lambda \sim \epsilon^{1/4}$ 
(Fig.~\ref{fig:wavelength}a), we conclude that 
both integrals in Eq.~(\ref{eq:lambda-1}) are dominated by the bulk of the wrinkled region rather than by the vicinity of its boundaries.








\subsection{How does the sheet's length affect the wavelength ? }\label{subsubsec:length}
We have seen above that the dependence of the wrinkle wavelength on the elastic moduli ($B$ and $Y$) and the exerted tensile load ($T$), expressed through the dimensionless parameters $\epsilon$ and $\tT$, agrees very well with the prediction of the CM model (Eqs.~\ref{eq:CM-1a},\ref{eq:CM-1a-dimensionless}). While our simulations do not allow us to test directly the dependence of $\lambda$ on $\tL$, we elaborate here on the rationale of the CM prediction $\lambda \sim \tL^{1/2}$, from the perspective of the FT analysis, 
and discuss the asymptotic limit $\tL \to \infty$ in the 
Hookean FT regime 
(assuming fixed values of $\tT \ll 1$ and $\epsilon \ll \epsilon_c$).

Recalling that the tensional stiffness $\Ktens$ in Eq.~(\ref{eq:lambda-1}) is a product of TFT and thus independent on $\epsilon$, and assuming that both
integrals in the denominator and the numerator are dominated by the bulk of the transversely-confined zone, 
we consider the 
``asymptotically long" limit, 
$\tL \gg 1$. 
We may envision (at least) two different scenarios for the outcome of TFT in this  limit: \\

$\bullet$ A spatially-uniform confinement: 
\begin{gather}
{\rm Scenario \ A} \nonumber \\ 
\ \ \PhiTFT (x)^2 \sim \tL^\alpha \nonumber \\ 
\PhiTFT'(x)^2 \sim \tL^{\alpha-2}     
\label{eq:op-A}
\end{gather}
with an exponent $\alpha>0$. \\

$\bullet$ A spatially-nonuniform confinement:
\begin{gather}
{\rm Scenario \ B:} \nonumber \\ 
\PhiTFT (x)^2 \sim 
\left\{ 
\begin{array}{cc}
C_0(\tT) \cdot f(\tilde{x}/W)  & \tilde{x} <C \cdot W \\
C_1(\tT)  &   \tilde{x} > C \cdot W
\end{array} \right. \nonumber  \\
\PhiTFT'(x)^2 \sim 
\left\{ 
\begin{array}{cc}
\tfrac{1}{W^2}C_0(\tT)f'(\tilde{x}/W)^2 & \tilde{x} <C\cdot W \\
0  &   \tilde{x} > C\cdot W
\end{array} \right.
\label{eq:op-B}
\end{gather}
where $\tilde{x} = |x \pm (\tfrac{1}{2}L-x^*)|$ is the distance from the end of the transversely-confined zone, $C$ is some constant, 
$C_0(\tT)$ and  $C_1(\tT)$ vanish as $\tT \to 0$, and $f(\xi)$ is some function such that $f'(\xi)^2$ is integrable as $\xi \to 0$.

The rationale underlying scenario A, which echos an assumption made in the CM model, is that the stretched sheet ``feels" the 
clamping at the short edges 
everywhere in the wrinkled zone, even though the sheet is arbitrarily long. The rationale underlying scenario B is that for $\tL \gg 1$, the confinement varies spatially only in a region close to the clamped edges, whose extent is indifferent to the length of the sheet (and hence must scale with the width $W$). 
While inspection of our numerical TFT solution (Fig.~\ref{fig:confinement-func}a) seems to support scenario A, we emphasize that we cannot rule out scenario B, or even more complicated scenarios, since our simulations do not explore sufficiently broad interval of values of $\tL$.      

Assuming scenario A, one readily note that Eq.~(\ref{eq:lambda-1}) yields 
$\Ktens \sim T/L^2$, whereas model B yields  $\Ktens \sim T /  L W$. Consequently, we find that: 
\begin{gather}
 {\rm Scenario \ A:} \ \ \lambda/W \sim \tL^{1/2} \nonumber  \\
 {\rm Scenario \ B:} \ \ \lambda/W \sim \tL^{1/4}
 \label{eq:scenarios}
\end{gather}
Once again we find that the nature of the confinement function $\PhiTFT^2(x)$, 
which is derived from the geometrically-nonlinear TFT and is 
strictly distinct from the planar state, may affect a noticeable departure from the prediction of the CM model. Numerical simulations of sufficiently long sheets will help to elucidate the length dependence of the wrinkle wavelength. \\

While the confinement function is derived from TFT, which totally ignores the bending rigidity of the sheet, 
the wrinkling of physical,  
highly bendable sheet ({\emph{i.e.}} $0<\epsilon \ll \epsilon_c$), cannot be described by any of the scenarios in Eq.~(\ref{eq:scenarios}) for arbitrarily long sheets. 
To see this, note that $\lambda$ is trivially bounded by the sheet width $W$. 
Thus, for any $\epsilon>0$ there exists a maximal length: 
\begin{gather}
 {\rm Scenario \ A:} \ \ \tL_{max}(\epsilon) \sim \epsilon^{-1/2}  \nonumber  \\
 {\rm Scenario \ B:} \ \ \tL_{max}(\epsilon) \sim \epsilon^{-1} 
 \label{eq:scenarios-1}
\end{gather}
such that for a sheet longer than $\tL_{max}(\epsilon) \cdot W$, 
the energetically-favorable deformation is no longer a parallel array of wrinkles 
that occupy most of the sheet. 
The nature of the deformation in such highly bendable but ``superlong'' sheets ($\epsilon \ll \epsilon_c, \tL >\tL (\epsilon)$)  
is an interesting question for future studies, even though it may not be easily accessible for experiments.

\section{Discussion} \label{sec:discussion}

\label{sec:discussion} 
\subsection{Phase diagram} \label{subsec:diagram}

\begin{figure}
\begin{tikzpicture}
    \node at (0,0) {\includegraphics[width=0.9\linewidth]{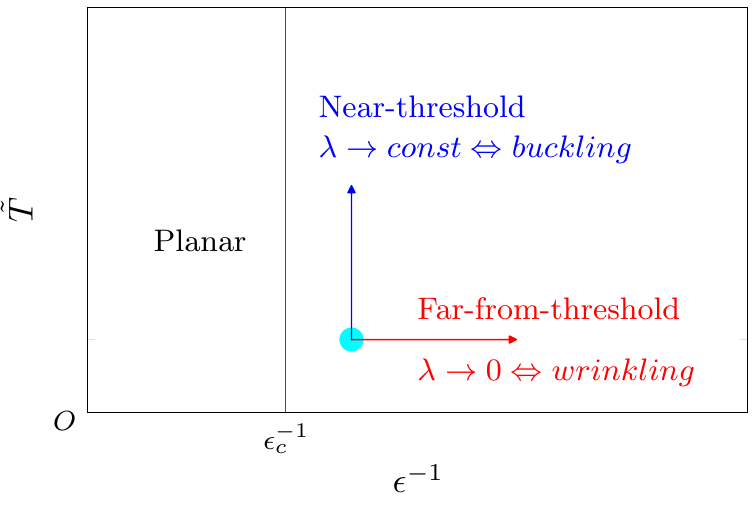}};
    \node at (0,-5.2) {\includegraphics[width=0.9\linewidth]{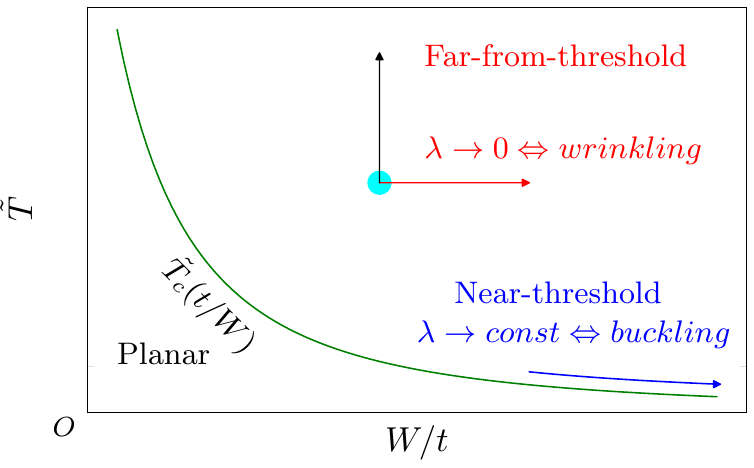}};
    \node at (-3.7, 2.39) {(A)};
    \node at (-3.7, -3.05) {(B)};
\end{tikzpicture}

\caption{
(A) A ``phase diagram" (for a fixed $\tL$), spanned by the bendability $\epsilon^{-1}$ and exerted tensile strain $\tT$. The threshold occurs at a vertical line, $\epsilon = \epsilon_c$ (Eq.~\ref{eq:Tc}). 
Close to the threshold line, $\epsilon \lesssim \epsilon_c$ (NT regime), the sheet exhibits a buckling pattern 
with a 
wavelength $\lambda \approx W/3$, localized in the compressed zones of the planar stress (see Ref.~\cite{Xin20}). 
For $\epsilon \ll \epsilon_c$ (FT regime), the 
pattern consists of wrinkles (wavelength $\lambda \sim W\epsilon^{1/2}$), which expands throughout the whole sheet.    
(B) Re-plotting the above phase diagram where the axes are now the normalized thickness $W/t$ and $\tT$, the threshold occurs at a curve 
$\tT_c \sim (t/W)^2$.  }
\label{fig:phase-diag}
\end{figure}

Figure~\ref{fig:phase-diag} delineates a schematic ``phase diagram'' of the stretched Hookean sheet, combining primary lessons 
from our analysis in Ref.~\cite{Xin20} and the current paper. Considering a given value of $\tL$, the diagram we plot in 
Fig.~\ref{fig:phase-diag}a is spanned by the two dimensionless parameters, $\epsilon^{-1}$ and $\tT$, Eq.~(\ref{eq:non-dim-1}).
Below the vertical threshold line, $\epsilon > \epsilon_c$, Eq.~(\ref{eq:Tc}), 
the planar state is stable. (In Fig.~\ref{fig:phase-diag}b we re-plot the same diagram using as two independent parameters $\tT$ and $W/t$, 
in which the threshold is a curve $\tT_c \sim (t/W)^2$).  
For $\epsilon \lesssim \epsilon_c$, 
our analysis in Ref.~\cite{Xin20} revealed that the deformation is characterized by a buckling mode, whose wavelength is independent on $\epsilon$ ($\lambda \approx W/3$), and whose spatial extent is limited to the transversely compressed zone of the planar state. Such a deformation is properly described by standard NT approach -- linear stability analysis and post-buckling methods. 
When $\epsilon \ll \epsilon_c$, 
the deformation becomes a wrinkle pattern which expands throughout most of the sheet, with a wavelength that vanishes as 
$\lambda \sim \epsilon^{1/4}$. While studies of other model systems revealed a pronounced variation of the deformation between distinct wrinkle patterns in the respective NT and FT regimes, the stretched rectangular sheet is 
exceptional, exhibiting a transition from a regular buckling mode to fully-developed wrinkle pattern. Notably, this dramatic morphological transition is driven by a minute energetic gain. This has been hinted already in 
Ref.~\cite{Xin20}, where we showed that the maximal transverse compression in the planar state is barely 
a half percentile of the exerted longitudinal tensile load. Figure~\ref{fig:energy-converge} shows that the energetic gain of the TFT limit (which provides a lower bound for the energy of the fully wrinkled state) may be a tiny fraction    
of the elastic energy of the corresponding (unstable) planar state.  \\

\subsection{Summary and open questions}
The main accomplishment of the current paper is a numerical demonstration that the fully-developed wrinkle pattern observed upon stretching a thin rectangular sheet is described by the FT framework (Sec.~\ref{sec:th-1}) -- a singular expansion of the Hookean elastic energy around the TFT solution of where the small parameter is the inverse bendability $\epsilon$ -- similarly to other problems in which such a description is amenable for analytic calculations. In addition to elucidating that the formation of wrinkles is governed by the collapse of transverse compressive stress, rather than transverse strain, 
our SE simulations further elucidate the geometrically-nonlinear nature of the fully-developed wrinkle pattern.
In Eq.~(\ref{eq:CM-1b}) of the CM model, the geometrical 
nonlinearity has been incorporated by invoking that the 
amplitude-wavelength ratio is independent on the bending modulus ({\emph{i.e.}} the dimensionless parameter $\epsilon$). However, our analysis 
shows that this $\epsilon$-independent ratio, given by the confinement function derived from the compression-free TFT solution, is itself a nonlinear function of the exerted tensile strain $\tT$ even for arbitrarily small $\tT$. This observation illuminates yet another subtle manifestation of the geometrical nonlinearity (Eq.~\ref{eq:trans-strain}) underlying FT analysis. \\


The analysis we presented here is based on numerical simulations, 
yielding numerous observations on the TFT solution ($\epsilon = 0$) and on the 
wrinkle pattern ($0<\epsilon \ll \epsilon_c$). One may wish to explain these observations by developing and analyzing a simplified, 
analytically-tractable model. For the benefit of a motivated reader, we close by highlighting some of these unexplained observations. \\

\subsubsection{TFT solution}

$\bullet$ We found that TFT yields a nonlinear dependence of macroscale features on the exerted tensile strain, 
most notably the transverse contraction of the planar projection, $\tDeltaTFT \sim \tT^{4/5}$. We interpreted this finding as a signature 
of the geometrically nonlinear nature of TFT, even though -- similarly to the planar state (which predicts linear dependence on $\tT$) -- it depends explicitly only on the in-plane displacement field. 

{\emph{Is it possible to obtain the exponent $\tfrac{4}{5}$ analytically ? }} \\      

$\bullet$ Our numerical solution of TFT is limited to a single length 
($\tL=8$), 
hence hampering our ability to make predictions for $\tL \gg 1$ even at a qualitative level ({\emph{e.g.}} discerning between scenarios A and B in Subsec.~\ref{subsubsec:length}).

{\emph{It is possible to predict the qualitative nature of the TFT solution for $\tL \!\!\gg\!\! 1$ without 
simulating long sheets ?}}  \\


$\bullet$ In our analysis we employed a semi-one-dimensional (1D) approach, whereby we extracted from simulations 
central features, 
such as the confinement function $\Phi^2(x)$ and the planar transverse contraction $\tDelta(x)$, by integrating over the width of the sheet. 
However, the 
observed wrinkle patterns (Fig.~\ref{fig:profile}) hint at nontrivial spatial structure of the TFT solution, whereby transverse confinement is restritced to the central half of the sheet. 

{\emph{What gives rise to 
an apparent ``half-width" rule ? }}


\subsubsection{Wrinkle pattern}

$\bullet$ In our SE simulations we found various power laws that characterize the convergence of the residual (transverse copressive) stress, as well as various macroscale features of the wrinkle pattern in physical, highly bendable sheets ($0<\!\!\epsilon \!\!\ll \!\!\epsilon_c$), to the respective TFT values. 

{\emph{Is it possible to obtain analytic expressions for the exponents in the    
power laws in Figs.~\ref{fig:stress-collapse}b-c, \ref{fig:ell-converge}, \ref{fig:energy-converge},\ref{fig:confinement-func}c ?}} \\


$\bullet$ Our semi-1D analysis falls short of accounting for the slowly-varying envelope that modulates the wrinkle amplitude 
along the transverse axis (Fig.~\ref{fig:profile}). Although amplitude modulations induced by geometrical frustration 
have been observed in more symmetric set-ups \cite{Tovkach20}, our problem appears to be different, since clamping the pulled edges violates 
transnational symmetry and may thus the cause of a non-periodic pattern. 

{\emph{Is it possible to predict how non-symmetric boundary conditions affect non-periodicity of wrinkle patterns ?}}   \\   

We thank F. Brau, E. Cerda, J. Chopin, P. Damman, A. Kudroli, and N. Menon for valuable discussions. 
This research was funded by the National Science Foundation under grant DMR 1822439. Simulations were performed in the computing cluster of Massachusetts Green High Performance Computing Center (MGHPCC).

\appendix


\section{TFT simulations} 
\label{sec:App-TFT}
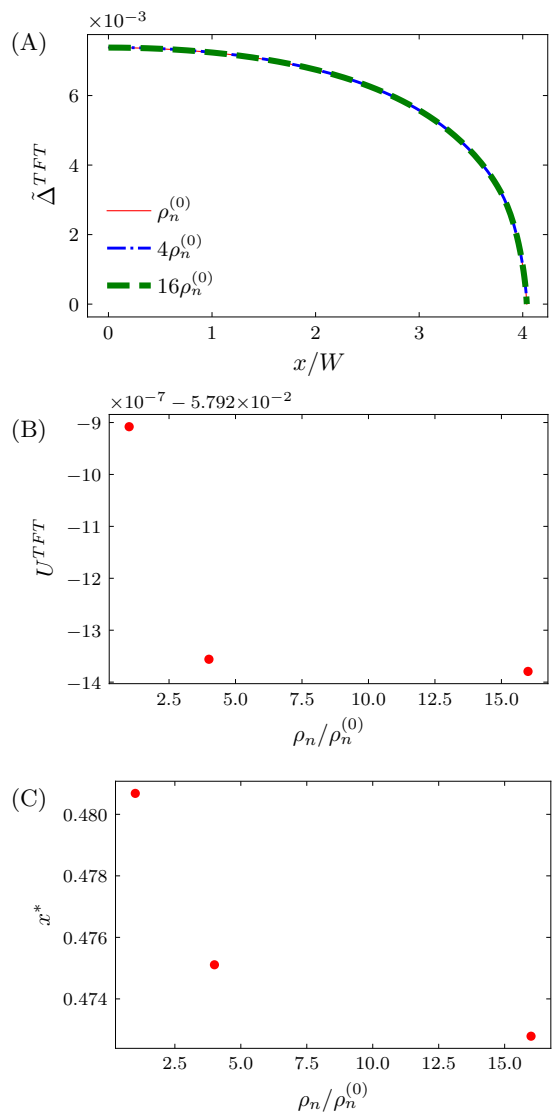
\begin{figure}
    \centering
    \begin{tikzpicture}
        \node at (0, 0) {\includegraphics[width=0.8\linewidth]{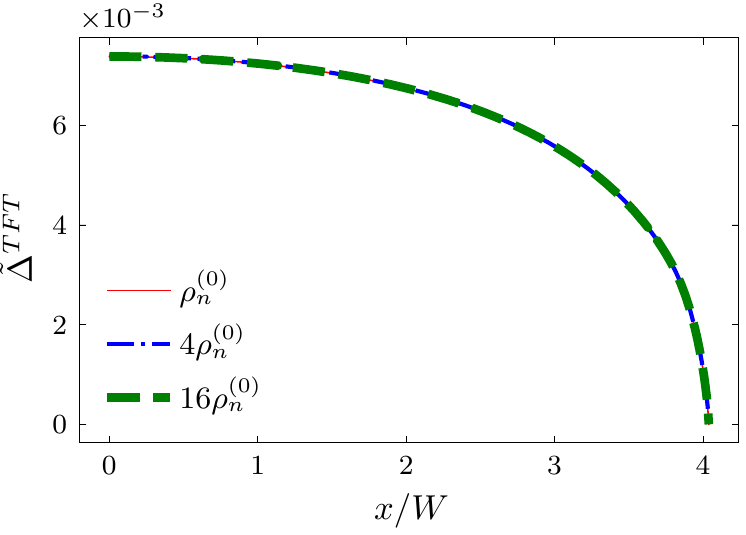}};
        \node at (0, -5.) {\includegraphics[width=0.8\linewidth]{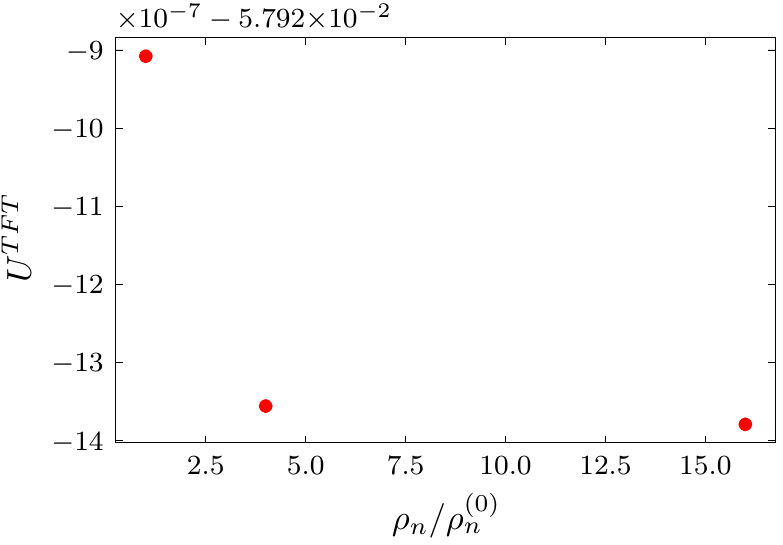}};
        \node at (0, -10) {\includegraphics[width=0.8\linewidth]{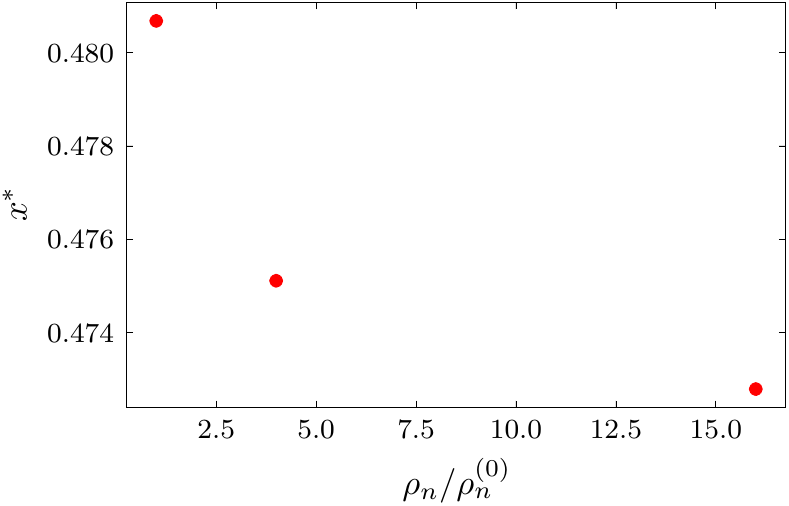}};
        \node at (-3.5, 2) {(A)};
        \node at (-3.5, -3.15) {(B)};
        \node at (-3.5, -8.05) {(C)};
    \end{tikzpicture}
    
    \caption{Various macroscale properties indicate convergence of our TFT simulations (a Hookean sheet with no bending rigidity) upon 
    increasing the mesh density $\rho_n$ beyond the standard value $\rho_n^{(0)} = 6.95 \times 10^5 $ used in most of the simulation: 
    (A) the transverse contraction $\tDeltaTFT(x)$, (B) the energy $U^{TFT}$, (C) the distance $x^*$ of the the transversely-contracted zone from the clamped edges. In both B and C, the differences ($|x^*(\rho_n) - x^*(\rho_n^{(0)}|$ and 
    $|U^{TFT}(\rho_n) - U^{TFT}(\rho_n^{(0)})|$, respectively) are much smaller than the corresponding differences from the values of the respective observables for sheets with finite bending rigidity ($0<\epsilon \ll 1$).}    
    \label{fig:tftconv}
\end{figure}

In simulating a sheet with no bending rigidity, any compression gives rise to an infinitely corrugated shape, limited only by the mesh 
size. In order to check that these simulations provide the TFT solution reliably, we performed simulations with a sequence of 
mesh densities, starting with the ``base'' density $\rho_n^{(0)} = 6.95 \times 10^5$, used in most of our simulations, then increasing the density to  
$4 \rho_n^{(0)}$ and to $16 \rho_n^{(0)}$. Figure~\ref{fig:tftconv} shows the numerical values of several macroscale features, 
which are predictable by TFT, for these mesh densities values. The variation among these different meshes is a tiny fraction ($\lesssim 10^{-3}$) of the 
characteristic differences between the TFT value and the finite-$\epsilon$ simulations, from which we extract the scaling laws in Figs.~\ref{fig:ell-converge},\ref{fig:energy-converge}).




\bibliography{MS-I}

\end{document}